\documentclass[aps, pra, reprint, twocolumn, superscriptaddress]{revtex4-2}

\usepackage{graphicx}
\usepackage{afterpage}
\usepackage{float}
\usepackage{amssymb, amsmath, amsfonts, amsopn, amstext, amsbsy}
\usepackage{bm}
\usepackage{color}
\usepackage{float}
\usepackage{setspace}
\usepackage{url}
\usepackage{latexsym}
\usepackage{epstopdf}
\usepackage{threeparttable}
\usepackage{cases}
\usepackage[colorlinks, linkcolor=blue, urlcolor=blue, citecolor=blue, pdfborder={0 0 0}, unicode]{hyperref}
\UseRawInputEncoding
\newcommand{\singlefigrefTwo}[1]{\hyperref[fig:2]{~2(#1)}}
\newcommand{\singlefigrefThree}[1]{\hyperref[fig:3]{~3(#1)}}
\newcommand{\singlefigrefFour}[1]{\hyperref[fig:4]{~4(#1)}}
\newcommand{\singlefigrefFive}[1]{\hyperref[fig:5]{~5(#1)}}
\newcommand{\singlefigrefSix}[1]{\hyperref[fig:6]{~6(#1)}}
\newcommand{\singlefigrefSeven}[1]{\hyperref[fig:7]{~7(#1)}}
\newcommand{\singlefigrefEight}[1]{\hyperref[fig:8]{~8(#1)}}
\newcommand{\singlefigrefNine}[1]{\hyperref[fig:9]{~9(#1)}}

\begin{document}
	
	\title{Nonreciprocal entanglement in exciton optomechanics \\ with an optical parametric amplifier}
	
	\author{Zhen-Sen Lin}\thanks{Co-first authors with equal contribution}
	\affiliation{Key Laboratory of Low-Dimensional Quantum Structures and Quantum Control of Ministry of Education, Hunan Research Center of the Basic Discipline for Quantum Effects and Quantum Technologies,  XJ-Laboratory and Department of Physics, Hunan Normal University, Changsha 410081, China}
	
	\author{Rui Zhang}\thanks{Co-first authors with equal contribution}

	\affiliation{Key Laboratory of Low-Dimensional Quantum Structures and Quantum Control of Ministry of Education, Hunan Research Center of the Basic Discipline for Quantum Effects and Quantum Technologies, XJ-Laboratory and Department of Physics, Hunan Normal University, Changsha 410081, China}
	
\author{Zi-Wei Jiang}
	\affiliation{Key Laboratory of Low-Dimensional Quantum Structures and Quantum Control of Ministry of Education, Hunan Research Center of the Basic Discipline for Quantum Effects and Quantum Technologies,  XJ-Laboratory and Department of Physics, Hunan Normal University, Changsha 410081, China}

\author{Wen-Quan Yang}
\affiliation{Key Laboratory of Low-Dimensional Quantum Structures and Quantum Control of Ministry of Education, Hunan Research Center of the Basic Discipline for Quantum Effects and Quantum Technologies,  XJ-Laboratory and Department of Physics, Hunan Normal University, Changsha 410081, China}

\author{Ya-Feng Jiao}\email{yfjiao@zzuli.edu.cn}
	\affiliation{Academy for Quantum Science and Technology, Zhengzhou University of Light Industry, Zhengzhou 450002, China}

\author{Hui Jing}\email{jinghui73@foxmail.com}
\affiliation{Key Laboratory of Low-Dimensional Quantum Structures and Quantum Control of Ministry of Education, Hunan Research Center of the Basic Discipline for Quantum Effects and Quantum Technologies,  XJ-Laboratory and Department of Physics, Hunan Normal University, Changsha 410081, China}
\affiliation{Academy for Quantum Science and Technology, Zhengzhou University of Light Industry, Zhengzhou 450002, China}

\author{Le-Man Kuang}\email{lmkuang@hunnu.edu.cn}
	\affiliation{Key Laboratory of Low-Dimensional Quantum Structures and Quantum Control of Ministry of Education, Hunan Research Center of the Basic Discipline for Quantum Effects and Quantum Technologies,  XJ-Laboratory and Department of Physics, Hunan Normal University, Changsha 410081, China}
	\affiliation{Academy for Quantum Science and Technology, Zhengzhou University of Light Industry, Zhengzhou 450002, China}
	


\begin{abstract}
		
We study nonreciprocal bipartite and tripartite entanglement in a spinning exciton-optomechanical system (EOMS)  with an optical parametric amplifier (OPA). We demonstrate that nonreciprocal  entanglement among photons, excitons, and phonons can be achieved under experimentally feasible parameters.
We find that the nonreciprocal  entanglement induced by Sagnac effects  can be regulated through the OPA.  Particularly, We show that the OPA significantly enhances photon-exciton entanglement and tripartite entanglement but weakens photon-phonon and exciton-phonon entanglement.  Moreover, we find that the photon-exciton nonreciprocal entanglement  not only can be generated at room temperature and even higher temperature but also exhibits  highly robustness to cavity dissipation.  Our works open a way to manipulate  the room-temperature nonreciprocal  entanglement, which may be useful for developing  nonreciprocal quantum technologies.
	
\end{abstract}

  \maketitle
\section{\label{level}Introduction}

Quantum entanglement \cite{1,2,3,4,5,6,7,8,9,10,11,12} describes a correlation between quantum mechanical systems, that does not occur in classical physics. It not only lies at the heart of quantum mechanics, but also is an essential quantum resource for emerging quantum information technologies \cite{13,14,15,16,17}, such as quantum computation, quantum communication, and quantum precise measurement. Cavity optomechanical systems  \cite{18,19,20,21} have become an important platform both for fundamental physics of macroscopic quantum systems  \cite{22,23,24,25,26,27,28} and for practical applications of precision sensing \cite{29,30,31,32}.
The cavity fields in cavity optomechanical systems exert radiation pressure on the movable mirror, which leads to the changes of both the resonance frequency and damping rate of the mechanical modes. At the same time, the mechanical vibration of the spring modulates the position of the movable mirror, which changes the cavity length and optical resonant frequency.
The resonant enhancement of both mechanical and optical response in the cavity optomechanical systems has enabled precision sensing of multiple physical quantities, including displacements, masses, forces, accelerations, magnetic fields, and ultrasounds.
Recently, owing to the remarkable progress achieved in ground-state cooling  \cite{33,34,35,36}, much attention has been paid to generate and manipulate quantum entanglement  in cavity optomechanical systems \cite{37,38,39,40,41,42,43,44,45,46,47,48,49}. Quantum entanglement between two oscillators has been achieved experimentally  \cite{50,51,52,53}.

An exciton optomechanical system (EOMS) is a new type of hybrid optomechanical system with exciton-photon and phonon-photon couplings. It has been experimentally
achieved in a GaAs/AlAs quantum well (QW) microcavity \cite{54,55,56,57,58}.  Microcavity exciton polaritons have many advantages \cite{54,55,59,60} such as strong light-matter couplings, low pump intensities, strong nonlinear effects, and high-quality factors.
Therefore, many novel and unique phenomena have been discovered in exciton polaritons such as room-temperature
polariton condensation \cite{61,62}, superfluidity \cite{63}, bistability \cite{64,65,66}, and optomechanically induced transparency \cite{67,68,69}. Meanwhile, exciton polaritons have also potential applications in designing advanced polaritonic devices such as polariton lasers \cite{70},  polariton spin switches and memories \cite{71,72},   polariton transistors \cite{55,73,74},  and polariton tunneling diodes \cite{75}.    In particular,  quantum entanglements in an  exciton optomechanical system  \cite{76} has also been studied. It has been shown that stationary exciton-photon or polariton-polariton entanglement can be achieved and  room temperature polariton entanglement can potentially be obtained by improving relevant experimental parameters of the EOMS \cite{76-1}.   Nevertheless, the possibility of achieving nonreciprocal bipartite and tripartite entanglement in the EOMS remains unexplored.

In this paper we investigate  nonreciprocal  entanglement in a spinning  EOMS with an optical parametric amplifier (OPA).  We demonstrate that nonreciprocal  entanglement among photons, excitons, and phonons can be achieved in this system under experimentally feasible parameters.
We find that the nonreciprocal  entanglement induced by Sagnac effects  can be manipulated  through the OPA.  Particularly, We show that the OPA significantly enhances photon-exciton entanglement and tripartite entanglement but weakens photon-phonon and exciton-phonon entanglement.  Moreover, we find that the photon-exciton nonreciprocal entanglement  is highly robust to thermal noise and cavity dissipation, retaining substantial values at room temperatures and even higher temperature.

This paper is structured as follows. In Sec.~\ref{level2}, we present the physical model and derive the linearized quantum Langevin equations of the system  under our consideration. In Sec.~\ref{level3} and Sec.~\ref{level4}, we study numerically  nonreciprocal bipartite and tripartite entanglement using experimentally accessible parameters, respectively.
Finally, concluding remarks are summarized in Sec.~\ref{level5}.

\section{\label{level2} Physical model and quantum Langevin equations}

\begin{figure}[htbp]
	\centering
	\includegraphics[width=0.56\textwidth]{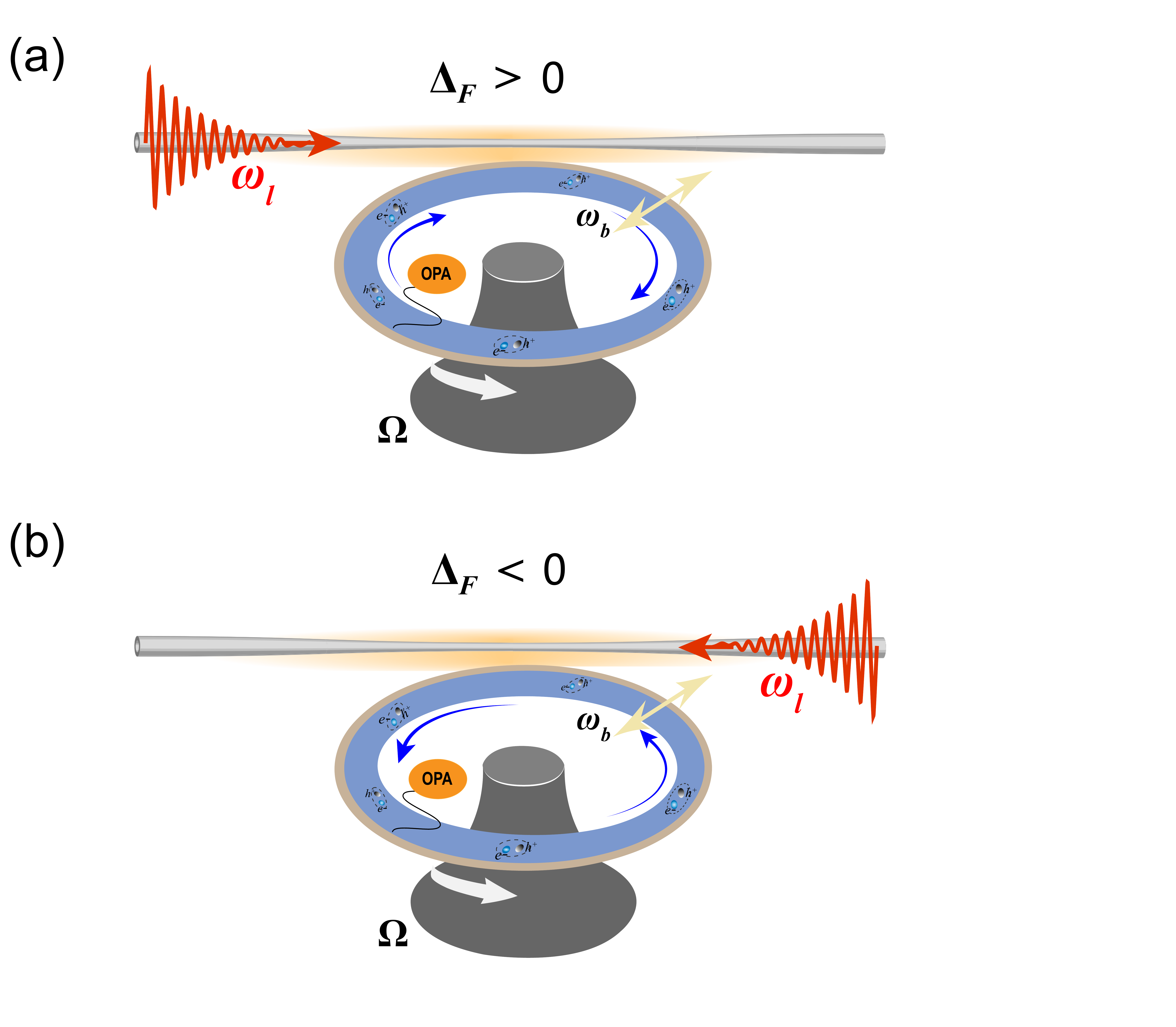}
	\caption{Schematic of a spinning exciton optomechanical system (EOMS) embeded an exciton-trapping quantum well (QW) and  an optical parametric amplifier (OPA). The resonator simultaneously supports a mechanical mode, an optical whispering-gallery mode, and an exciton mode.  The microwave driving can be applied respectively along two different directions: $\Delta_F > 0$ denotes the driving fields comes from the left-hand side  (a), $\Delta_F < 0$ denotes the driving field  comes from the right-hand side  (b).}
	\label{fig:1}
\end{figure}

 In this section, we propose the physical model under our consideration, present the derivation of the linearized quantum Langevin equations (QLEs) and subsequently calculate the steady-state solution of the system. We consider an exciton optomechanical system with a spinning microresonator and an optical parametric amplifier as shown in Fig.~\ref{fig:1} in which a rotating whispering-gallery microresonator is evanescently  coupled with a tapered fiber. The microresonator includes InGaAs quantum wells (QWs) for trapping excitons. A radiation-pressure induces a mechanical radial breathing mode for acting as phonons.  An optical parametric amplifier (OPA) is embedded in an optomechanical cavity.  QWs are formed by multiple thick layers of InGaAs and embedded between two layers of AlGaAs. It is well known that due to the rotation, the optical mode frequency experiences Sagnac-Fizeau shift, which transforms as
\begin{align}
	\omega_{c} &\rightarrow \omega_{c} + \Delta_{F}, \\
	\Delta_{F} &= \left( \frac{n R \Omega \omega_{c}}{c} \right) \left( 1 - \frac{1}{n^{2}} - \frac{\lambda}{n} \frac{\mathrm{d} n}{\mathrm{d} \lambda} \right),
\end{align}
where $\Omega$ is the angular velocity of the spinning resonator; $n$ and $R$ are the refractive index and radius of the resonator, respectively; and $c$ and $\lambda$ are the speed of light and the light wavelength in a vacuum, respectively. The dispersion term $dn/d\lambda$ represents a negligibly small relativistic (dispersion) correction in the Sagnac-Fizeau shift. Here, we fix the clockwise rotation of the resonator, hence $\Delta_F > 0$ ($\Delta_F < 0$) corresponds to the situation of driving the resonator from its left-hand (right-hand) side, as shown in Fig.~\ref{fig:1}.


In a frame rotating at the drive frequency $\omega_{\text{0}}$, the Hamiltonian of the system reads ($\hbar = 1$):
\begin{equation}
	\begin{split}
		H =&\ (\Delta_c + \Delta_F) c^\dagger c + \Delta_a a^\dagger a + \omega_b b^\dagger b \\
		&+ J (c^\dagger a + a^\dagger c) + g c^\dagger c (b^\dagger + b) \\
		&+ iG (c^{\dagger 2} e^{i\varphi} - c^{2} e^{-i\varphi}) + i\varepsilon (c^\dagger - c),
	\end{split}
\end{equation}
where $c$, $a$ and $b$ ($c^{\dagger}$, $a^{\dagger}$ and $b^{\dagger}$) are the annihilation (creation) operators of the cavity photon, exciton and phonon, respectively, satisfying the commutation relation $[j, j^\dagger] = 1$ ($j = c,a,b$),  and $\omega_c$, $\omega_a$, and $\omega_b$ are their resonance frequencies. The detuning of the cavity frequency $\omega_{c}$ from the drive field is $\Delta_c = \omega_{c} - \omega_{0}$, the exciton mode frequency $\omega_{a}$ is $\Delta_a = \omega_{a} - \omega_{0}$. The  Rabi coupling strength between cavity photons and QW excitons is  $J$. The coupling effect between cavity photons and phonons is described by a single-photon optomechanical coupling strength $g = \omega_c / R \sqrt{\hbar / (2 m_{\text{eff}} \omega_b)}$, with $m_{\text{eff}}$ the effective mass of mechanical modes. The single-photon optomechanical coupling strength $g$ is typically weak, but the effective optomechanical coupling can be significantly enhanced by driving the microcavity with an intense laser field. The parameter $G$ represents the nonlinear gain of the OPA with $\varphi$ as the phase of the driving field. The last term corresponds to the single-photon drive, where $\varepsilon = \sqrt{2P \kappa_c / \hbar \omega_0}$ signifies the coupling strength between the cavity and the drive field with frequency $\omega_0$, power $P$ and cavity decay rates $\kappa_c$.

By incorporating the dissipation and input noise of each mode, we obtain the following quantum Langevin equations (QLEs):
\begin{equation}
	\label{eq:QLEs}
	\begin{aligned}
		\dot{c} =& -[i(\Delta_{c} + \Delta_{F}) + \kappa_{c}] c - i J a - i g c (b^{\dagger} + b)\\
		&+ 2 G c^{\dagger} e^{i\phi} + \varepsilon + \sqrt{2 \kappa_{c}} c^{\text{in}}, \\
		\dot{a} =& -(i \Delta_{a} + \kappa_{a}) a - i J c + \sqrt{2 \kappa_{a}} a^{\text{in}}, \\
		\dot{b} =& -(i \omega_{b} + \kappa_{b}) b - i g c^{\dagger} c + \sqrt{2 \kappa_{b}} b^{\text{in}},
	\end{aligned}
\end{equation}
where  $\kappa_{c}$, $\kappa_{a}$, $\kappa_{b}$ represent the decay rates of the optical, excitonic, and mechanical modes, respectively.  $j^{\text{in}}(t)$ $(j = c, a, b)$ are the input noise operators of the three modes, which are zero-mean and characterized by the correlation functions $\langle j^{\text{in}}(t) j^{\text{in}\dagger}(t') \rangle = [N_{j}(\omega_{j}) + 1] \delta(t-t')$ and $\langle j^{\text{in}\dagger}(t) j^{\text{in}}(t') \rangle = N_{j}(\omega_{j}) \delta(t-t')$, with $N_{j}(\omega_{j}) = 1 / (e^{\hbar \omega_{j} / k_{B} T} - 1)$ being the equilibrium mean thermal excitation number of mode $j$, and $T$ the bath temperature.

\vspace{-\parskip}
We can linearize the dynamics of the system if we strongly drive the cavity photon mode such that $\lvert \langle c \rangle \rvert \gg 1$. Similarly, a strong exciton-photon excitation-exchange interaction, leads to $\lvert \langle a \rangle \rvert \gg 1$. Therefore, we can write an operator in terms of its steady-state value and a fluctuation part up to first order: $j = \langle j \rangle + \delta j$, where $\langle j \rangle$ is the steady-state mean value, and $\delta j$ is a small fluctuation operator with zero mean value. Inserting the operator forms in Eq.\eqref{eq:QLEs}, the QLEs are separated into two sets of equations, one for the steady-state average values and the other for the quantum fluctuations. The linearized QLEs for the quantum fluctuations are obtained as
\begin{equation}
	\label{eq:linearized_QLEs}
	\begin{aligned}
		\dot{\delta c} =& -(i\widetilde{\Delta }_c + i\Delta_F + \kappa_c)\delta c - iJ\delta a - G_{cb}(\delta b^\dagger + \delta b) \\
		&\quad + 2 G e^{i\varphi} \delta c^\dagger + \sqrt{2\kappa_c} c^{\text{in}}, \\
		\dot{\delta a} =& -(i\Delta_a + \kappa_a)\delta a - iJ\delta c + \sqrt{2\kappa_a} a^{\text{in}}, \\
		\dot{\delta b} =& -(i\omega_b + \kappa_b)\delta b +G_{cb}^*\delta c - G_{cb} \delta c^\dagger + \sqrt{2\kappa_b} b^{\text{in}},
	\end{aligned}
\end{equation}
where $\widetilde{\Delta }_c = \Delta_c + 2g\langle b\rangle$ is the effective cavity-drive detuning including the frequency shift due to the optomechanical interaction, and $G_{cb} = i g \langle c \rangle$ is the effective optomechanical coupling strength. The expressions of the steady-state average values  are given by
\begin{equation}
	\label{eq:steady_state}
	\begin{aligned}
		\langle c \rangle = &\varepsilon \frac{2G e^{i\varphi} + \varLambda}{|\varLambda|^2 - 4G^2}, \\
		\langle a \rangle = &-\frac{iJ}{i\Delta_a + \kappa_a} \langle c \rangle, \\
		\langle b \rangle = &-\frac{g}{\omega_b} |\langle c \rangle|^2,
	\end{aligned}
\end{equation}
where the complex coefficients $\varLambda = -i(\widetilde{\Delta }_c + \Delta_F) + \kappa_c + {J^2}/(-i\Delta_a +\kappa_a)$.

\vspace{-\parskip}
The QLEs \eqref{eq:linearized_QLEs} can be expressed in a compact matrix form with the quadratures of fluctuation
and noise operators $\delta X_j = (\delta j + \delta j^\dagger)/\sqrt{2}$, $\delta Y_j = i(\delta j^\dagger - \delta j)/\sqrt{2}$ and $X_j^{\text{in}} = (j^{\text{in}} + j^{\dagger{\text{in}}})/\sqrt{2}$, $Y_j^{\text{in}} = i(j^{\dagger{\text{in}}} - j^{\text{in}})/\sqrt{2}$, i.e.,
\begin{equation}
	\label{eq:matrix_form}
	\dot{\mathbf{u}}(t) = \mathbf{A} \mathbf{u}(t) + \mathbf{n}(t),
\end{equation}
where $\mathbf{u}(t) = [\delta X_c(t), \delta Y_c (t), \delta X_a (t), \delta Y_a (t), \delta X_b(t),\delta Y_b(t)]^{\mathrm{T}}$ represents the vector of quantum fluctuations, $\mathbf{n}(t) = [\sqrt{2\kappa_c}X_c^{\text{in}},\sqrt{2\kappa_c}Y_c^{\text{in}},\sqrt{2\kappa_a}X_a^{\text{in}},\sqrt{2\kappa_a}Y_a^{\text{in}},\sqrt{2\kappa_b}X_b^{\text{in}}, \\
\sqrt{2\kappa_b}Y_b^{\text{in}}]^{\mathrm{T}}$ is the vector of input noises, and the drift matrix $\mathbf{A}$ is given by

\begin{equation}
	\mathbf{A} =
	\begin{pmatrix}
		-\kappa_{\varphi,-} & \tilde{\Delta}_{\varphi,+} & 0 & J & -2\operatorname{Re}G_{cb} & 0 \\
		-\tilde{\Delta}_{\varphi,-} & -\kappa_{\varphi,+} & -J & 0 & -2\operatorname{Im}G_{cb} & 0 \\
		0 & J & -\kappa_a & \Delta_a & 0 & 0 \\
		-J & 0 & -\Delta_a & -\kappa_a & 0 & 0 \\
		0 & 0 & 0 & 0 & -\kappa_b & \omega_b \\
		-2\operatorname{Im}G_{cb} & 2\operatorname{Re}G_{cb} & 0 & 0 & -\omega_b & -\kappa_b
	\end{pmatrix},
	\label{eq:drift_matrix_A}
\end{equation}
where $\tilde{\Delta}_{\varphi,\pm} = \tilde{\Delta}_{c} + \Delta_{F} \pm 2G\sin\varphi$, $\kappa_{\varphi,\pm} = \kappa_{c} \pm2G\cos\varphi$. The solution of the above compact form equation can be given directly, $\mathbf{u}(t) = \mathbf{M}(t) \mathbf{u}(0) + \int_{0}^{t} ds\, \mathbf{M}(s) \mathbf{n}(t-s)$, where $\mathbf{M}(t) = \exp(\mathbf{A}t)$. According to the Routh-Hurwitz criterion, the system eventually reaches stability only when all eigenvalues of $\mathbf{A}$ have the negative real parts. In our proposed parameter scheme, the criterion can be satisfied, then we can have $\mathbf{M}(\infty)=0$ in the steady state and
\begin{equation}
	\mathbf{u}_i(\infty) = \int_0^\infty ds \underset{k}{\sum} \mathbf{M}_{ik}(s)\mathbf{n}_k(t-s).
	\label{eq:steady_state_solution}
\end{equation}

Due to the linearized dynamics and the Gaussian nature of the quantum noises, the steady state of the system, independently of any initial conditions, finally evolves into a tripartite zeromean Gaussian state, which is fully characterized by a $6\times6$ correlation matrix (CM) $\mathbf{V}$, with its components
\begin{equation}
	\mathbf{V}_{ij}=\langle \mathbf{u}_i(\infty)\mathbf{u}_j(\infty)+\mathbf{u}_j(\infty)\mathbf{u}_i(\infty)\rangle/2.
	\label{eq:correlation_matrix}
\end{equation}

By substituting Eq.\eqref{eq:steady_state_solution} into Eq.\eqref{eq:correlation_matrix} and using the fact that the six components of $\mathbf{n}(t)$ are uncorrelated, the steady-state correlation matrix $\mathbf{V}$ is obtained as
\begin{equation}
	\mathbf{V} = \int_{0}^{\infty} ds\, \mathbf{M}(s) \mathbf{D} \mathbf{M}^{\mathrm{T}}(s),
	\label{eq:integral_representation}
\end{equation}
where $\mathbf{D} = \mathrm{Diag}[\kappa_{c}(2N_{c}+1),\kappa_{c}(2N_{c}+1),\kappa_{a}(2N_{a}+1),\kappa_{a}(2N_{a}+1),\kappa_{b}(2N_{b}+1),\kappa_{b}(2N_{b}+1)]$ is the  diffusion matrix, defined through $\langle \mathbf{n}_i(s) \mathbf{n}_j(s') + \mathbf{n}_j(s') \mathbf{n}_i(s) \rangle /2 = \mathbf{D}_{ij} \delta(s - s')$. Under the stability condition, the steady state correlation matrix $\mathbf{V}$ fulfills the Lyapunov equation
\begin{equation}
	\mathbf{A}\mathbf{V} + \mathbf{V}\mathbf{A}^{\mathrm{T}} = -\mathbf{D}.
	\label{eq:lyapunov_equation}
\end{equation}
Clearly, it is seen that the Lyapunov equation \eqref{eq:lyapunov_equation} is linear for $\mathbf{V}$ and can be solved straightforwardly, thereby implying that one can derive the CM $\mathbf{V}$ for any specific value of relevant parameters.

\section{\label{level3}  Nonreciprocal bipartite  entanglement}

In this section, we study the generation and control of nonreciprocal bipartite entanglement among  the optical mode, the excitonic mode, and the mechanical mode.
We adopt the logarithmic negativity as a quantitative measure of the quantum entanglement between any two modes of the system, which is defined as
\begin{equation}
	E_{N} = \max\!\big[\,0,\, -\ln(2\nu^{-}) \,\big]
	\label{eq:log_negativity}
\end{equation}
where $\nu^- = \min[ \lvert \operatorname{eig}(i\Omega_2\tilde{\mathbf{V}}_4)\rvert]$ (the symplectic matrix $\Omega_{2} = \oplus_{j=1}^{2}i\sigma_{y}$ and $\sigma_{y}$ is the y-Pauli matrix) is the minimum symplectic eigenvalue of $\tilde{\mathbf{V}}_4 = P \mathbf{V}_4 P$, with $\mathbf{V}_4$ being the reduced CM of the two modes under consideration, obtained by removing in $\mathbf{V}$ the rows and columns associated with the uninteresting modes, and $P = \mathrm{Diag}[1,-1,1,1]$ being the matrix that performs partial transposition on the reduced CM. Eq.\eqref{eq:log_negativity} quantifies how much the positivity of the partial transpose condition for separability is violated for the considered Gaussian states, which is equivalent to Simon’s necessary and sufficient entanglement nonpositive partial transpose criterion (or the related Peres-Horodecki criterion). Note that the selected two modes of  system get entangled if and only if $\nu^- < 1/2$, where $E_N$ has a nonzero value.

\begin{figure}[htbp]
	\centering
	\includegraphics[width=0.5\textwidth]{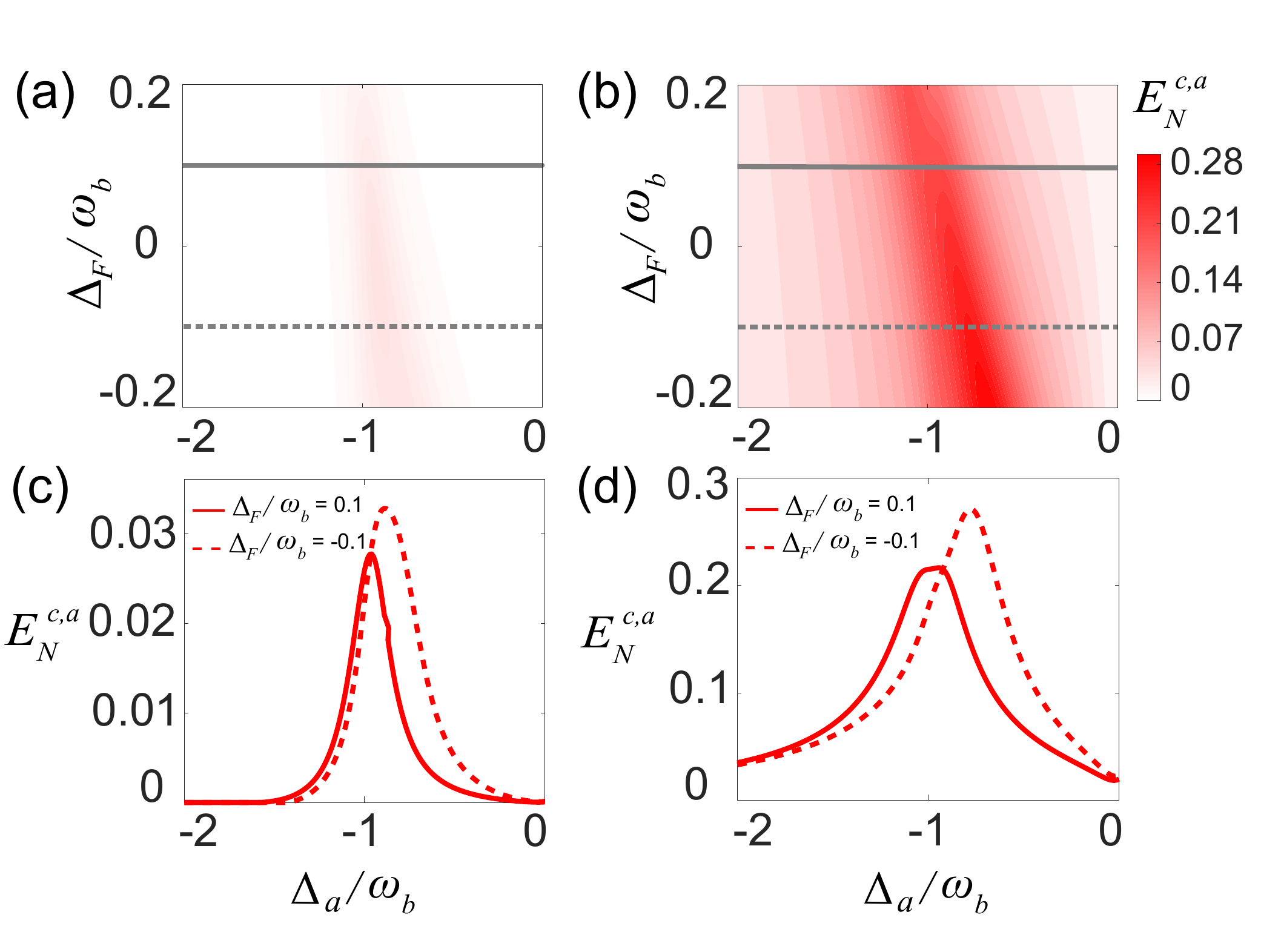}
	\caption{The photon-exciton entanglement ($E_{N}^{c,a}$) versus $\Delta_{F}/\omega_{b}$ and the exciton-drive detuning $\Delta_{a}/\omega_{b}$. (a) $G = 0$, (b) $G = 0.08\omega_{b}$, the phase $\varphi$ associated with the OPA drive field is assumed to be $0$. Moreover, we set $\Delta_{F}/\omega_{b} = 0.1$ (solid line) and $\Delta_{F}/\omega_{b} = -0.1$ (dashed line) in (c)-(d). We take $\tilde{\Delta}_{c} = 0.9\omega_{b}$, and the other parameters are provided in the text.}
	\label{fig:2}
\end{figure}

In the following, we numerically investigate nonreciprocal bipartite  entanglement in the exciton optomechanical system under our consideration. We show that nonreciprocal bipartite  entanglement  can be manipulated via the OPA.
In our calculations, we have selected experimentally feasible parameter values: $\omega_{b}/2\pi$ = 1GHz, $\omega_{0}/2\pi$ = 345THz, $\kappa_{c}/2\pi$ = $\kappa_{a}/2\pi$ = 80MHz, $\kappa_{b}/2\pi$ = 100KHz, $J/2\pi$ = 280MHz, $g/2\pi$ = 500KHz, $\varepsilon/2\pi$ = 0.1THz, and at temperature $T$ = 10mK. These chosen values of parameters  are in stable regions of the system under our consideration according to the Routh-Hurwitz criterion.
Firstly, we study bipartite entanglement between optical and exciton modes.
In Fig.~\ref{fig:2}, we plot the bipartite entanglement ($E_{N}^{c,a}$) between optical and exciton modes as a function of $\Delta_{F}/\omega_{b}$ and the exciton-drive detuning $\Delta_{a}/\omega_{b}$ in the absence as well as the presence of the OPA. Fig.\singlefigrefTwo{a} and Fig.\singlefigrefTwo{b} show the change of entanglement with respect to parameters $\Delta_{F}/\omega_{b}$ and $\Delta_{a}/\omega_{b}$  when the OPA is absent and present, respectively. By comparing Fig.\singlefigrefTwo{a} and Fig.\singlefigrefTwo{b}, it can be found that in the presence of the OPA, a significant enhancement in $E_{N}^{c,a}$ is observed. From Fig.\singlefigrefTwo{b} we can see that entanglement is enhanced roughly by one order of magnitude in the presence of the OPA, and the parameter region where entanglement survives is also significantly expanded.
It is worth noting that the presence of the OPA is crucial for the enhancement of entanglement in this system. The main reason is that the OPA enhances the cavity occupation number and changes the photon statistics of the cavity mode thereby improving cavity-exciton coupling. As a result, the steady-state value of the cavity mode [see Eq.\eqref{eq:steady_state}] also depends on parametric gain $G$ and associated phase $\varphi$.
In Fig.\singlefigrefTwo{c} and\singlefigrefTwo{d}, we plot the photon-exciton entanglement with respect to the detuning $\Delta_{a}/\omega_{b}$  without and  with the OPA for  $\Delta_{F}/\omega_{b}=\pm 0.1$, respectively. Comparing Fig.\singlefigrefTwo{c} with Fig.\singlefigrefTwo{d}, we can find that  the OPA can not only  significantly enhance the amount of photon-exciton entanglement but also the  nonreciprocity of entanglement.

\begin{figure}[htbp]
	\centering
	\includegraphics[width=0.5\textwidth]{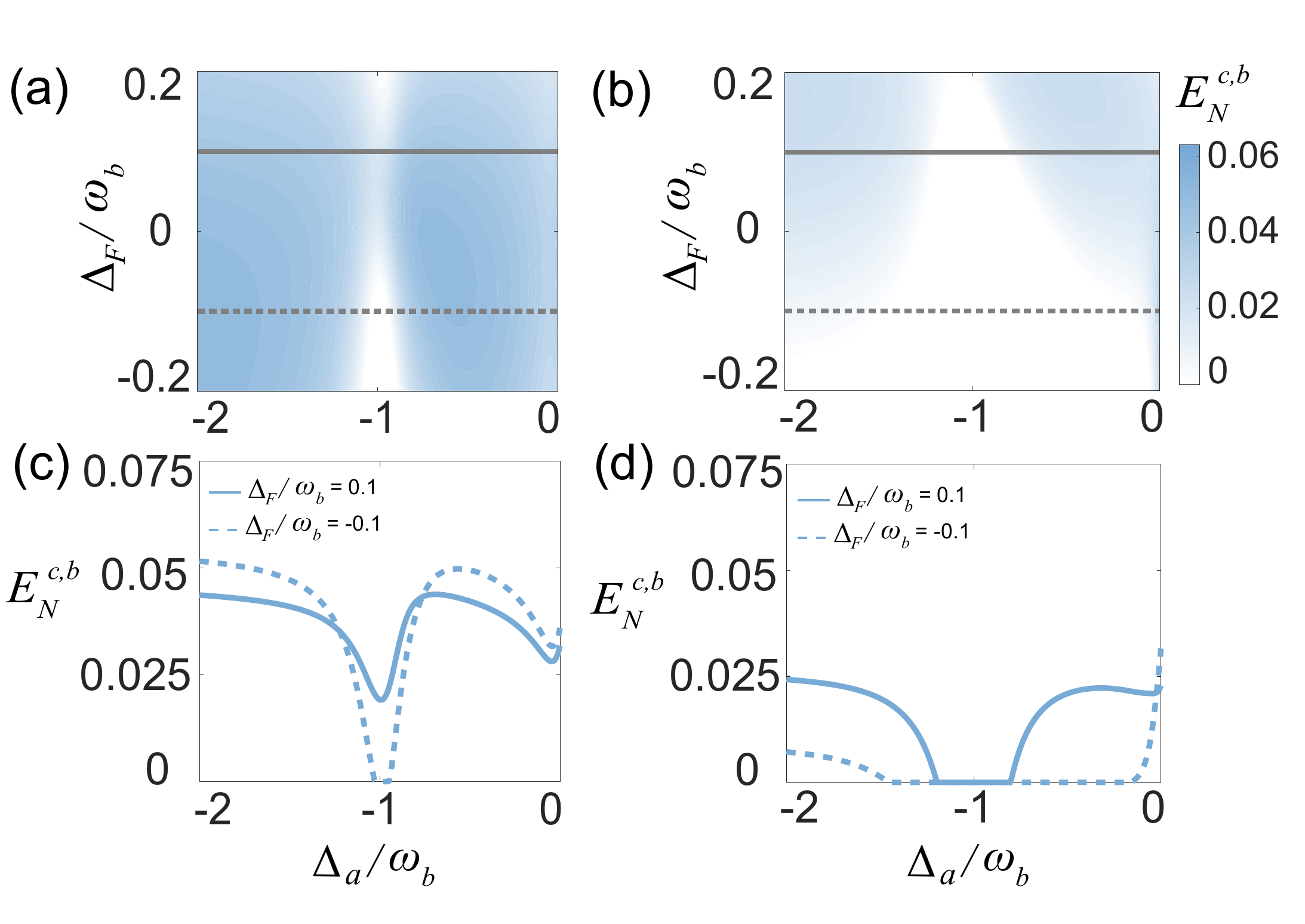}
	\caption{The photon-phonon  ($E_{N}^{c,b}$) versus $\Delta_{F}/\omega_{b}$ and the exciton-drive detuning $\Delta_{a}/\omega_{b}$ (a) $G = 0$, (b) $G = 0.08\omega_{b}$, the phase $\varphi$ associated with the OPA drive field is assumed to be $\pi/2$. Moreover, we set $\Delta_{F}/\omega_{b} = 0.1$ (solid line) and $\Delta_{F}/\omega_{b} = -0.1$ (dashed line) in (c)-(d). The other parameters are the same as in Fig.~\ref{fig:2}.}
	\label{fig:3}
\end{figure}

We then study the bipartite entanglement  between optical and mechanical modes. We plot the photon-phonon entanglement ($E_{N}^{c,b}$)   as a function of $\Delta_{F}/\omega_{b}$ and the exciton-drive detuning $\Delta_{a}/\omega_{b}$  in Fig.~\ref{fig:3}. Here we take the phase of the OPA drive field to be $\pi/2$ while the values of other parameters are the same as in Fig.~\ref{fig:2}.  Different from the situation of the photon-exciton entanglement, from Fig.\singlefigrefThree{a} and Fig.\singlefigrefThree{b} we can observe that the OPA   generally  weakens  the photon-phonon entanglement.  In particular, in the region of $\Delta_{F}/\omega_{b} <0.1$, photon-phonon entanglement is almost completely suppressed.  Interestingly, when $\Delta_{a}/\omega_{b} = -1$, we observe an ideal nonreciprocity: the cavity-phonon entanglement can be generated for $\Delta_{F} > 0$, whereas no entanglement occurs for $\Delta_{F} < 0$ [see Fig.\singlefigrefThree{c}]. It is remarkable that the OPA extends the range of ideal nonreciprocity over a broader detuning parameter space compared to the system without OPA, particularly at phase $\varphi = \pi/2$, as demonstrated in Fig.\singlefigrefThree{d}. In Fig.\singlefigrefThree{c} and\singlefigrefThree{d}, we plot the photon-phonon entanglement with respect to the detuning $\Delta_{a}/\omega_{b}$  without and  with the OPA for  $\Delta_{F}/\omega_{b}=\pm 0.1$, respectively. From  Fig.\singlefigrefThree{c} and Fig.\singlefigrefThree{d}, we can see that  the OPA weakens photon-phonon entanglement but enhances its nonreciprocity.

\begin{figure}[htbp]
	\centering
	\includegraphics[width=0.5\textwidth]{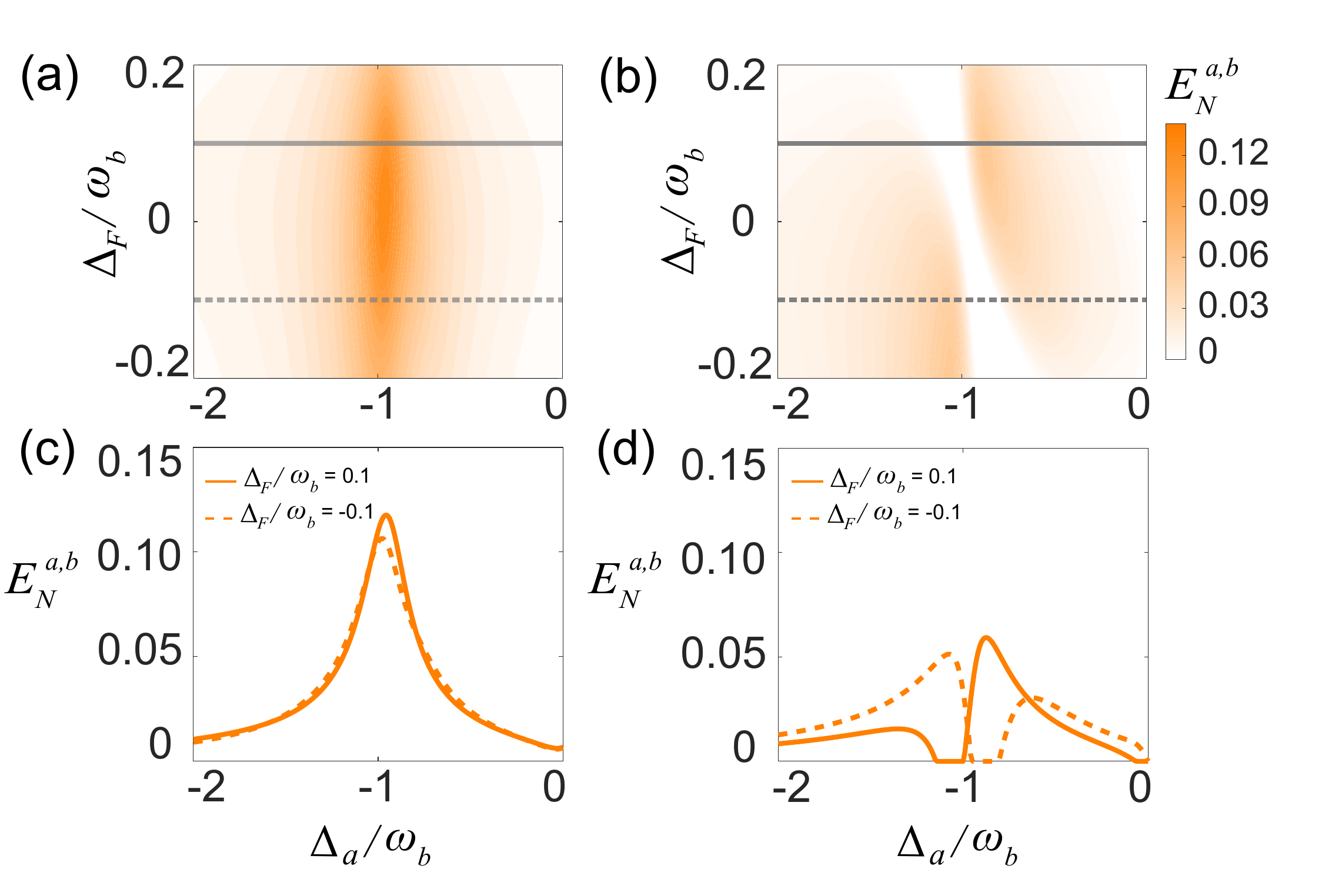}
	\caption{The exciton-phonon entanglement ($E_{N}^{a,b}$) versus $\Delta_{F}/\omega_{b}$ and the exciton-drive detuning $\Delta_{a}/\omega_{b}$ (a) $G = 0$, (b) $G = 0.08\omega_{b}$, the phase $\varphi$ associated with the OPA drive field is assumed to be $0$. Moreover, we set $\Delta_{F}/\omega_{b} = 0.1$ (solid line) and $\Delta_{F}/\omega_{b} = -0.1$ (dashed line) in (c)-(d). The other parameters are the same as in Fig.~\ref{fig:2}.}
	\label{fig:4}
\end{figure}

We now turn to the bipartite entanglement  between exciton and mechanical modes. In this case, the exciton-phonon entanglement ($E_{N}^{a,b}$) as a function of $\Delta_{F}/\omega_{b}$ and the exciton-drive detuning $\Delta_{a}/\omega_{b}$  is plotted in Fig.~\ref{fig:4}. Here the values of the relevant parameters are the same as in Fig.~\ref{fig:2}. Similar to the situation of the photon-phonon entanglement, from Fig.\singlefigrefFour{a} and Fig.\singlefigrefFour{b} we can observe that the OPA  not only fails to enhance the exciton-phonon entanglement, but also weakens it.  However, the presence of the OPA can lead to the stronger nonreciprocity of the exciton-phonon entanglement. In Fig.\singlefigrefFour{c} and\singlefigrefFour{d}, we plot the exciton-phonon entanglement with respect to the detuning $\Delta_{a}/\omega_{b}$. From Fig.\singlefigrefFour{d}, we can find that almost ideal nonreciprocal entanglement can emerge  in the nearby regime of $\Delta_{a}/\omega_{b}=-1$.

From Fig.~\ref{fig:2}, Fig.~\ref{fig:3} and Fig.~\ref{fig:4}, we can also observe that the bipartite entanglement among optical,  exciton, and mechanical modes is nonreciprocal. When the rotation direction of the microresonator remains unchanged,  the applied driving fields from left (right)-hand side induces a positive (negative) frequency shift $\Delta_{F}$ due to the Sagnac effect. Thus, nonreciprocal entanglement can be generated by the Sagnac effect in this system. 
In order to  characterize nonreciprocal entanglement quantitatively, we assess the nonreciprocal degree of bipartite entanglements by the use of the bidirectional contrast ratio  defined by
    \begin{equation}
    \label{eq:contrast_E}
	\begin{aligned}
        C_{N}^{i,j}=&\frac{\left|E_{N}^{i,j}\left(\Delta_{F}>0\right)-E_{N}^{i,j}\left(\Delta_{F}<0\right)\right|}{E_{N}^{i,j}\left(\Delta_{F}>0\right)+E_{N}^{i,j}\left(\Delta_{F}<0\right)},
	\end{aligned}
    \end{equation}
which implies $0\leqslant C_{N}^{i,j} \leqslant1$. When $C_{N}^{i,j} = 1$, the system exhibits ideal nonreciprocity while $C_{N}^{i,j} = 0$ indicates a complete absence of nonreciprocity for bipartite  entanglements. A greater value of contrast ratio $C_{N}^{i,j}$ signifies a stronger nonreciprocity of entanglement. 

\begin{figure*}[htbp]
	\centering
	\includegraphics[width=0.85\textwidth]{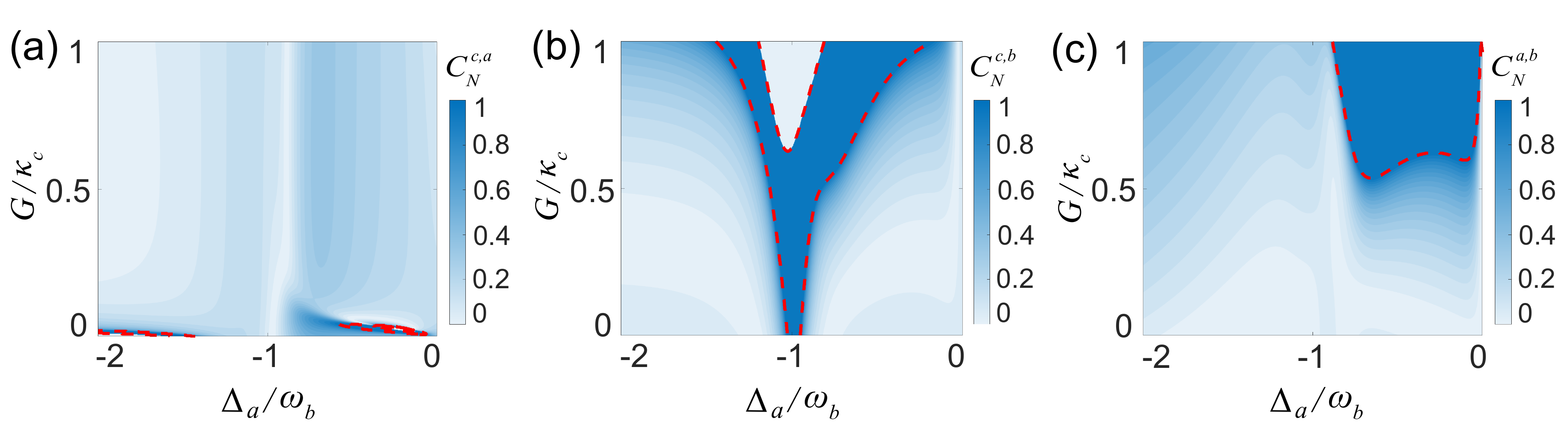}
	\caption{Bidirectional contrast ratios (a) $C_{N}^{c,a}$, (b) $C_{N}^{c,b}$ and (c) $C_{N}^{a,b}$ versus the OPA nonlinear gain $G/\kappa_{c}$ and the exciton-drive detuning $\Delta_{a}/\omega_{b}$, with phase $\varphi = \pi/2$ and $|\Delta_F| = 0.1\omega_b$. The region enclosed by the red dashed line indicates the ideal nonreciprocal entanglement. The other parameters are the same as in Fig.~\ref{fig:2}.}
	\label{fig:5}
\end{figure*}

To clearly observe the effect of OPA on the bidirectional contrast ratio $C_{N}^{i,j}$ of the bipartite nonreciprocal entanglement, we present in Fig.~\ref{fig:5} the dependence of $C_{N}^{i,j}$ on the OPA nonlinear gain $G/\kappa_{c}$ and the exciton-drive detuning $\Delta_{a}/\omega_{b}$ for phase $\varphi = \pi/2$ and $|\Delta_F| = 0.1\omega_b$. The results  clearly indicate that the ideal nonreciprocity of bipartite entanglements ($C_{N}^{c,a} = C_{N}^{c,b} = C_{N}^{a,b} = 1$) can be realized via control of the OPA nonlinear gain and the exciton-drive detuning, as indicated by the region enclosed by the red dashed lines. We can observe from Fig.\singlefigrefFive{a} that the ideal nonreciprocal region ($C_N^{c,a}=1$) for the photon-exciton entanglement appears in the range of small OPA gain ($G/\kappa_c \lesssim 0.1$) and is approximately symmetrically distributed on both sides of the detuning $\Delta_a/\omega_b = -1$. However, further increasing $G/\kappa_c$ beyond this range suppresses the nonreciprocal entanglement, leading to a degradation of $C_N^{c,a}$. In Fig.\singlefigrefFive{b} the OPA enables a substantial extension of the operational detuning range for achieving ideal nonreciprocal entanglement between the optical and  mechanical modes. The ideal nonreciprocal region ($C_N^{c,b}=1$) expands  around $\Delta_a/\omega_b = -1$ as the OPA gain increases, covering a progressively wider detuning interval. This behavior indicates that the OPA not only facilitates but also robustly maintains the nonreciprocity of photon-phonon entanglement over a broad parameter space. Remarkably, from Fig.\singlefigrefFive{c} we can see that in the absence of OPA or when the OPA nonlinear gain is very small ($G/\kappa_{c}\ll1$), only weak nonreciprocity ($C_{N}^{a,b}\approx0$) emerges in the exciton-phonon entanglement. However, within the detuning interval $\Delta_a/\omega_b \in [-1,0]$, ideal nonreciprocity in the exciton-phonon entanglement is achieved when $G/\kappa_{c}\approx0.5$ and extends over a wider parameter space with increasing $G/\kappa_{c}$.

\begin{figure}[htbp]
	\centering
	\includegraphics[width=0.47\textwidth]{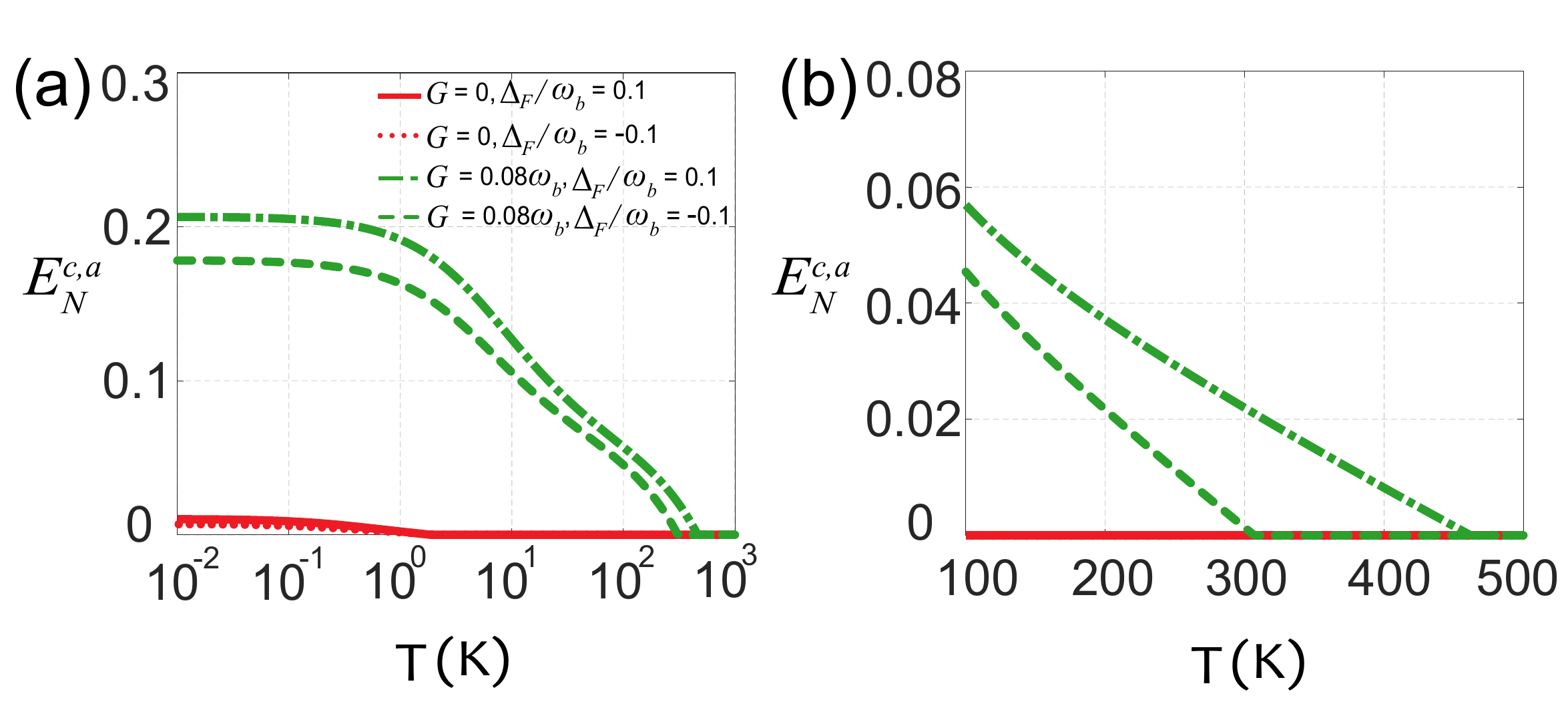}
	\caption{(a) Nonreciprocal photon-exciton entanglement $E_N^{c,a}$ as a function of temperature $T$ in the presence ($G = 0.08\omega_b$, $\varphi = 0$) and the absence ($G = 0$) of the OPA; (b) magnified view of the same data to highlight the behavior near the vanishing point with the presence of the OPA. The parameters are $\tilde{\Delta}_c = \omega_b$, $\Delta_a = -\omega_b$, $g/2\pi = 300\,\mathrm{KHz}$, and the other parameters are the same as in Fig.~\ref{fig:5}.}
	\label{fig:6}
\end{figure}

In what follows we discuss the influence of temperature on the bipartite entanglement. In Fig.\singlefigrefSix{a} we plot the photon-exciton entanglement as a function of temperature in the absence and presence of the OPA. The results reveal that the presence of the OPA significantly enhances the robustness of entanglement against temperature. This implies that the strong tolerance of the generated photon-exciton entanglement against thermal noise. The red solid and dotted curve in Fig.\singlefigrefSix{a} show that entanglement $E_{N}^{c,a}$ vanishes at a temperature of around 2 K in the absence of the OPA. At this same temperature, the green dash-dot (dashed) curve, which is plotted for $\varphi = 0$ in the presence of the OPA, shows an entanglement value of $E_{N}^{c,a}$ approximately 0.18 (0.15). Therefore, the presence of the OPA facilitates strong entanglement at a much higher temperature. Fig.\singlefigrefSix{b} is  the magnified view  of  Fig.\singlefigrefSix{a} between 100K and 500K. It highlights the temperature dependence of $E_N^{c,a}$ near the vanishing point of entanglement in the presence of the OPA. Notably, it clearly shows that the entanglement persists up to  much higher temperatures 450 K for $\Delta_{F}/\omega_{b} = 0.1$ and 300 K for $\Delta_{F}/\omega_{b} = -0.1$. The results further demonstrate that the OPA can significantly enhance the thermal robustness of the photon-exciton entanglement, and reveal the possibility of achieving room-temperature (300 K) and even high-temperature photon-exciton entanglement through parameter optimization. 
				
\begin{figure}[htbp]
	\centering
	\includegraphics[width=0.5\textwidth]{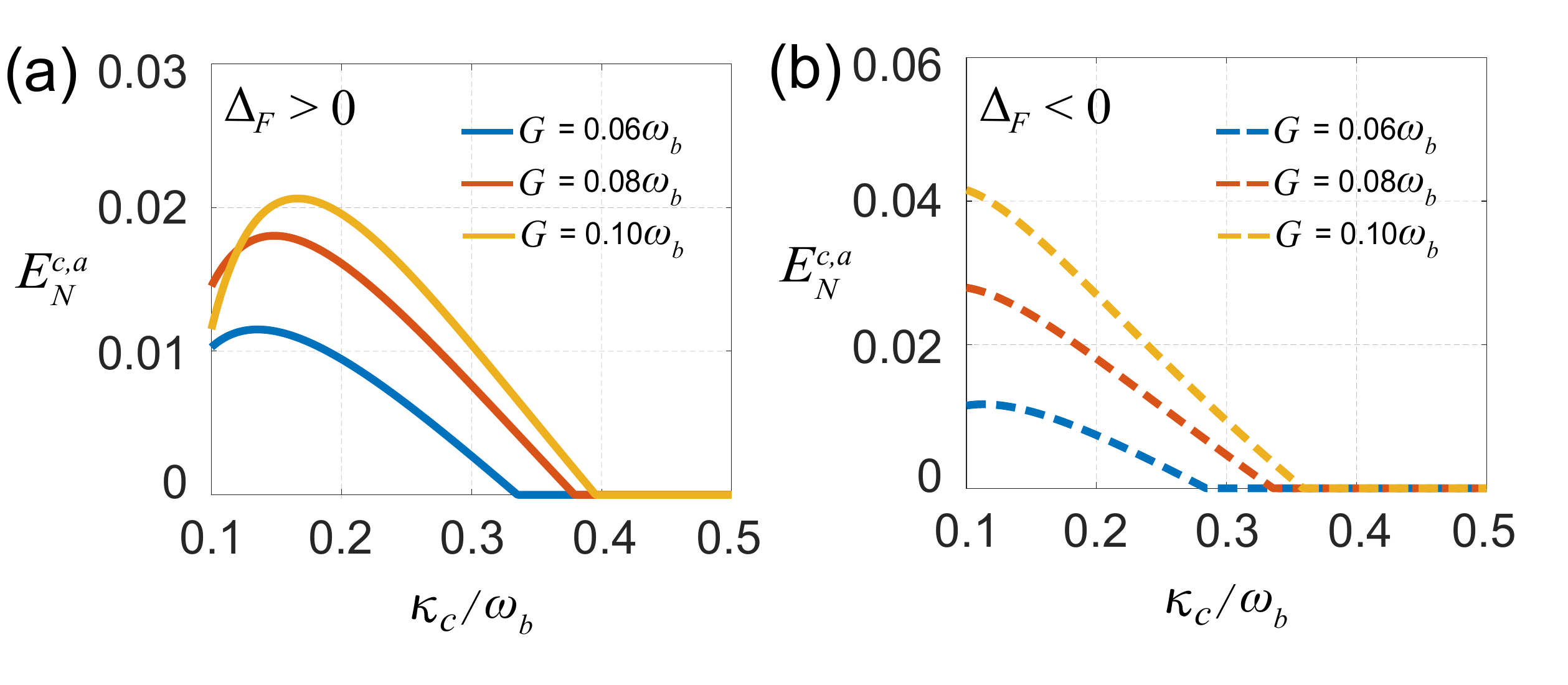}
	\caption{Nonreciprocal photon-exciton entanglement $E_N^{c,a}$ as a function of cavity decay rate $\kappa_{c}/\omega_{b}$ in the presence of the OPA with different nonlinear gains G for (a) $\Delta_{F}/\omega_{b} = 0.1$ and (b) $\Delta_{F}/\omega_{b} = -0.1$. We take temperatures $T = 260 K$, and the other parameters are the same as in Fig.~\ref{fig:6}.}
	\label{fig:7}
\end{figure}

We now investigate the robustness of the photon-exciton entanglement against cavity dissipation under different OPA gains. Figure~\ref{fig:7} shows the photon-exciton entanglement $E_N^{c,a}$ as a function of the cavity decay rate $\kappa_c/\omega_b$ for $\Delta_F/\omega_b = 0.1$ (a) and $\Delta_F/\omega_b = -0.1$ (b), with the OPA nonlinear gain set to $G = 0.06\omega_b$, $0.08\omega_b$, and $0.10\omega_b$, at a fixed temperature $T = 260\,\mathrm{K}$. The other parameters are the same as in Fig.~\ref{fig:6}.
From Fig.\singlefigrefSeven{a} we can see that for each $G$ value, $E_N^{c,a}$ initially increases with $\kappa_c/\omega_b$, reaching a maximum at an optimal cavity decay rate, and then gradually decreases. Notably, as the OPA gain $G$ increases, the decay of entanglement becomes significantly slower, and the range of $\kappa_c$ over which non-zero entanglement persists expands considerably. This indicates that a larger OPA gain effectively enhances the resilience of the photon-exciton entanglement against cavity dissipation. A similar trend is observed in Fig.\singlefigrefSeven{b} for $\Delta_F/\omega_b = -0.1$, where higher $G$ again leads to stronger robustness against dissipation. These results demonstrate the positive role of the OPA in improving the noise resistance of entanglement, offering a pathway to achieve robust quantum entanglement in highly dissipative environments.

\begin{figure}[htbp]
	\centering
	\includegraphics[width=0.48\textwidth]{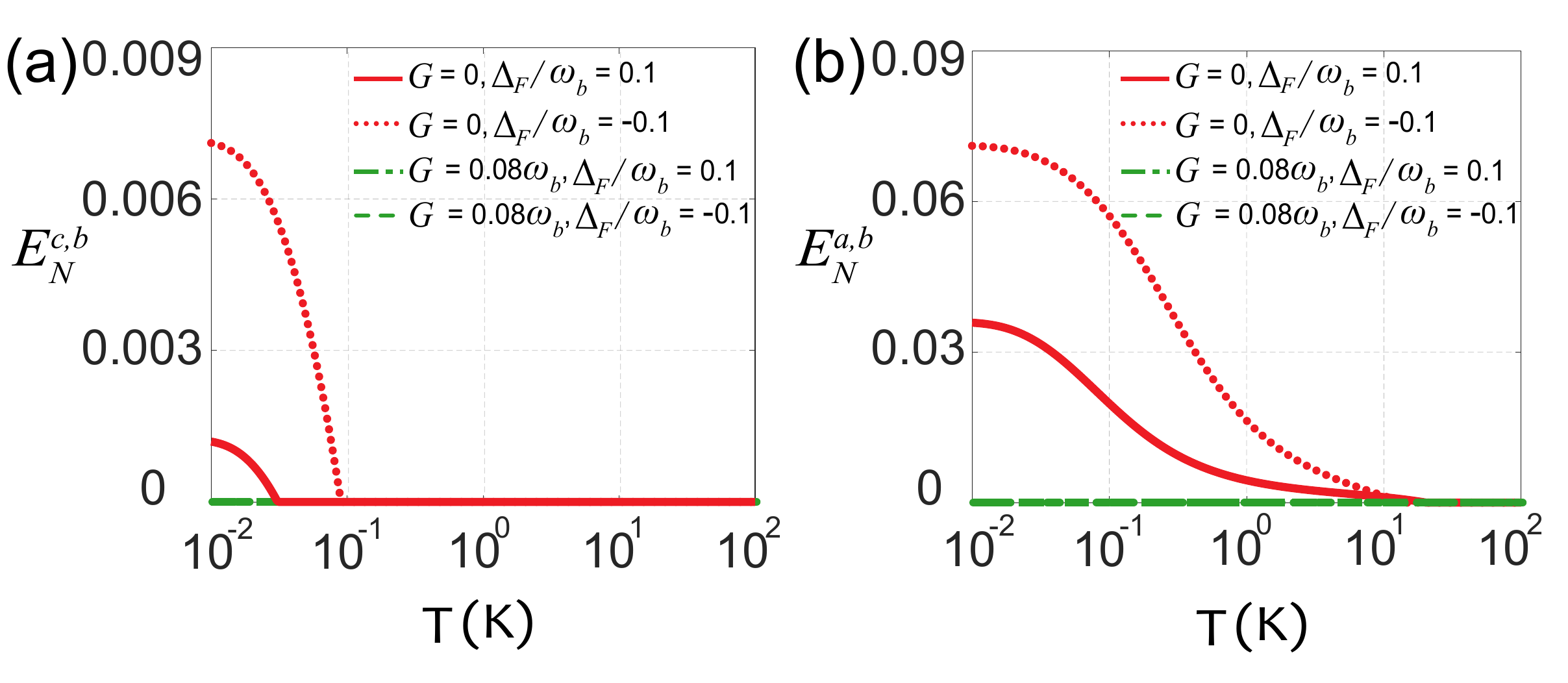}
	\caption{(a) Nonreciprocal photon-phonon entanglement $E_{N}^{c,b}$ and (b) exciton-phonon entanglement $E_{N}^{a,b}$ as a function of temperature $T$ in the presence ($G = 0.08\omega_{b}$, $\varphi = 0$) and the absence ($G = 0$) of the OPA. The other parameters are the same as in Fig.~\ref{fig:6}.}
	\label{fig:8}
\end{figure}

It is worth noting that photon-phonon and exciton-phonon entanglements exhibit weaker resistance against thermal noise than  photon-exciton entanglement. 
In Fig.~\ref{fig:8}, we plot photon-phonon and exciton-phonon entanglements as a function of temperature, respectively. From Fig.\singlefigrefEight{a} and \singlefigrefEight{b}, we can see that  both the photon-phonon and exciton-phonon entanglement become more vulnerable to thermal effects in the presence of the OPA, with  entanglement-vanishing temperatures considerably reduced. The results suggest that the OPA-mediated nonlinearity couples differently to the mechanical mode, potentially introducing additional thermal noise channels that accelerate the thermal degradation of these phonon-mediated entanglements.

\section{\label{level4}  nonreciprocal tripartite entanglement}

In this section, we study tripartite entanglement among the optical, exciton and mechanical modes. We adopt the minimum residual contangle as a quantitative measure of tripartite entanglement \cite{77,78}.
It is defined by
\begin{equation}
	R_{\tau}^{\min} = \min\!\big[ R_{\tau}^{\,i|jk},\ R_{\tau}^{\,j|ik},\ R_{\tau}^{\,k|ij} \big],
	\label{eq:min_residual_tangle}
\end{equation}
where residual entanglement is given by
\begin{equation}
	R_{\tau}^{\,i|jk} = C_{i|jk} - C_{i|j} - C_{i|k},
	\label{eq:residual_tangle_definition}
\end{equation}
where $C_{u|v}$ is the contangle of subsystems of $u$ and $v$ ($v$ contains one or two modes), which is a proper entanglement monotone defined as the squared logarithmic negativity. To calculate the one-mode-vs-two-modes logarithmic
negativity $E_{i|jk}$, one only needs to follow the definition of Eq.\eqref{eq:log_negativity} simply by replacing $\Omega_{2} = \oplus_{j=1}^{2}i\sigma_{y}$ with $\Omega_{3} = \oplus_{j=1}^{3}i\sigma_{y}$, and $\tilde{\mathbf{V}}_4 = P \mathbf{V}_4 P$ with $\tilde{\mathbf{V}} = P_{i|jk} \mathbf{V} P_{i|jk}$, where $P_{1|23} = \mathrm{Diag}[1,-1,1,1,1,1]$, $P_{2|13} = \mathrm{Diag}[1,1,1,-1,1,1]$ and $P_{3|12} = \mathrm{Diag}[1,1,1,1,1,-1]$ are partial transposition matrices.


\begin{figure}[htbp]
	\centering
	\includegraphics[width=0.5\textwidth]{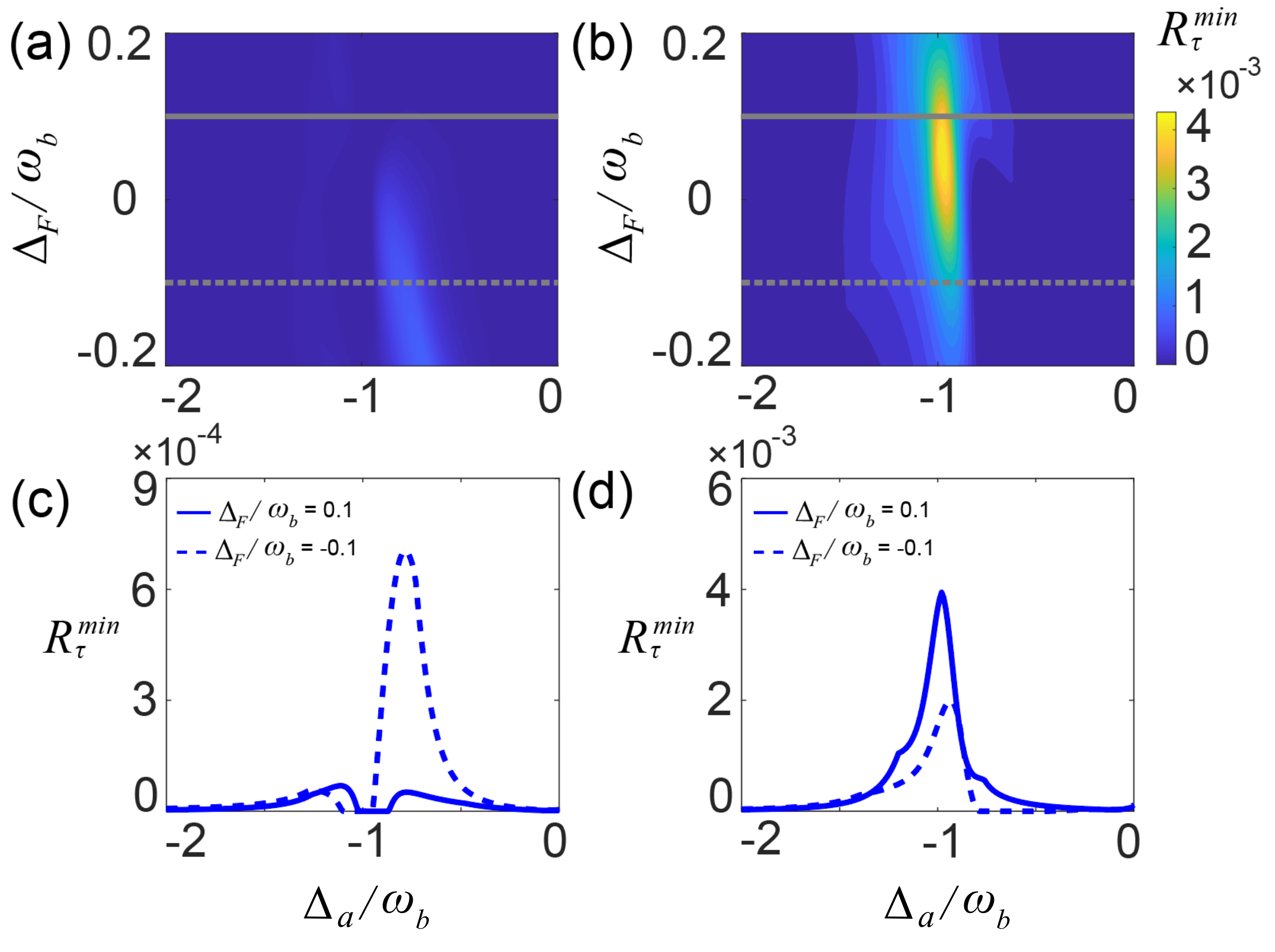}
	\caption{The tripartite entanglement between cavity-exciton-phonon modes ($R_{\tau}^{min}$) versus $\Delta_{F}/\omega_{b}$ and the exciton-drive detuning $\Delta_{a}/\omega_{b}$ (a)$G = 0$, (b)$G = 0.08\omega_{b}$, $\varphi = \pi/2$. We set $\Delta_{F}/\omega_{b} = 0.1$ (solid line) and $\Delta_{F}/\omega_{b} = -0.1$ (dashed line) in (c)-(d). The other parameters are the same as in Fig.~\ref{fig:2}.}
	\label{fig:9}
\end{figure}

In Fig.~\ref{fig:9}, we plot the minimum residual contangle $R_{\tau}^{min}$ versus $\Delta_{F}/\omega_{b}$ and the detuning $\Delta_{a}/\omega_{b}$ in the absence as well as the presence of the OPA. From  Fig.\singlefigrefNine{a} and Fig.\singlefigrefNine{b} we can observe that the presence of the OPA leads to a significant enhancement of the tripartite entanglement.  Fig. \singlefigrefNine{b} indicates that the optimal parameter region is given by $-0.05<\Delta_{F}/\omega_{b} <0.14$ and $-1.4<\Delta_{a}/\omega_{b}<-0.8$. Comparing  Fig.\singlefigrefNine{c} and\singlefigrefNine{d} we can find that the maximum $R_{\tau}^{min}$ is increased  by nearly an order of magnitude  compared to the situation  without the OPA. This is an important feature which makes the experimental generation and detection of tripartite entanglement more feasible. 
Moreover, we can see that the strength of tripartite entanglement is notably lower than that of bipartite entanglement through comparing Fig.~\ref{fig:9} with Fig.~\ref{fig:2}, Fig.~\ref{fig:3}, and Fig.~\ref{fig:4}.


The quantum nonreciprocity of tripartite entanglement can be measured by the bidirectional contrast ratio defined by
\begin{equation}
	\label{eq:contrast_E}
	\begin{aligned}
		C_{R}=&\frac{\left|R_{\tau}^{min}\left(\Delta_{F}>0\right)-R_{\tau}^{min}\left(\Delta_{F}<0\right)\right|}{R_{\tau}^{min}\left(\Delta_{F}>0\right)+R_{\tau}^{min}\left(\Delta_{F}<0\right)},
	\end{aligned}
\end{equation}
which indicates that when $C_R= 1$, the system exhibits ideal nonreciprocal tripartite entanglement while $C_R = 0$ denotes a complete absence of nonreciprocity for tripartite  entanglement. A greater value of contrast ratio $C_R$ signifies a stronger nonreciprocity of  tripartite entanglement.

\begin{figure}[htbp]
	\centering
	\includegraphics[width=0.5\textwidth]{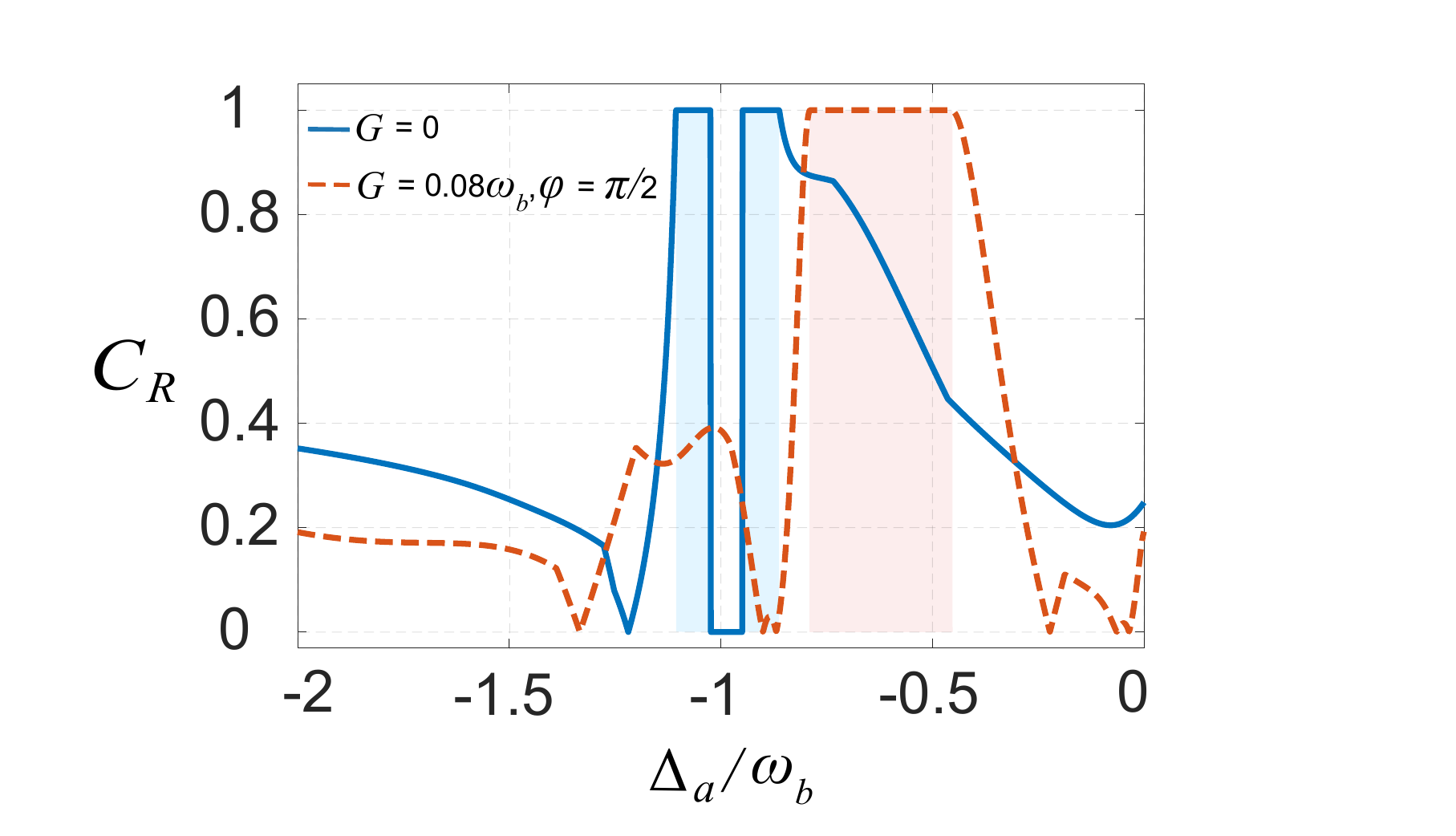}
	\caption{Bidirectional contrast ratio $C_{R}$ versus the exciton-drive detuning $\Delta_{a}/\omega_{b}$. The coloured vertical stripes represent the ideal nonreciprocal entanglement zones. The parameters are the same as in Fig.~\ref{fig:8}.}
	\label{fig:10}
\end{figure}

In Fig.~\ref{fig:10}, we plot the bidirectional contrast ratio $C_{R}$ versus  the detuning $\Delta_{a}/\omega_{b}$ in the absence as well as the presence of the OPA for $\Delta_F = \pm 0.1\omega_b$， which correspods to the drive field being applied from either the left or right side, respectively. The red dashed (blue solid) curve shows the bidirectional contrast ratio in the presence (absence) of the OPA. The results reveal that the presence of the OPA can change  the bidirectional contrast ratio of the tripartite entanglement.  In particular, we can observe that  the ideal nonreciprocal tripartite entanglement ($C_{R}=1$) exhibits a blue shift and a broadening of its operational range in the presence of the OPA, as indicated by the red shaded vertical stripes.


\begin{figure}[htbp]
	\centering
	\includegraphics[width=0.45\textwidth]{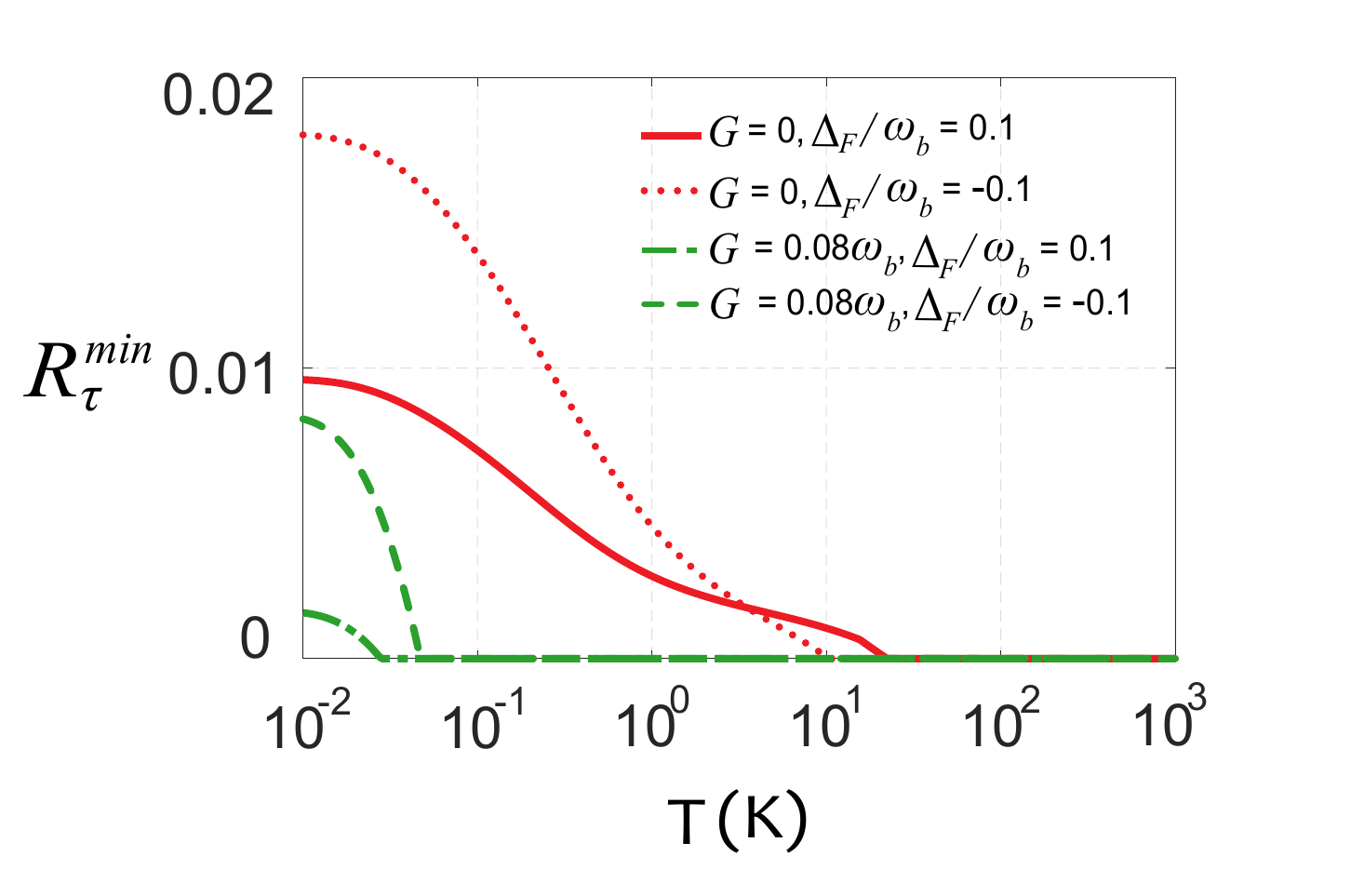}
	\caption{Nonreciprocal tripartite entanglement $R_{\tau}^{min}$ as a function of temperature $T$ in the presence ($G = 0.08\omega_b$, $\varphi = 0$) and the absence ($G = 0$) of the OPA. The other parameters are the same as in Fig.~\ref{fig:6}.}
	\label{fig:11}
\end{figure}

Finally, we investigate the effect of temperature on the generated nonreciprocal tripartite entanglement. In  Fig.~\ref{fig:11}, we plot the minimum residual contangle  $R_{\tau}^{min}$ as a function of temperature $T$ in the presence ($G = 0.08\omega_b$, $\varphi = 0$) and the absence ($G = 0$) of the OPA. The results show that the OPA significantly weakens the tolerance of the nonreciprocal tripartite entanglement against thermal noise, and reduces the amount of entanglement.  From Fig.~\ref{fig:11}, we can observe that temperature of living environment in the absence of the OPA can reach tens of K. It is two orders of magnitude higher than  that of the presence of the OPA. It is worth noting that the effect of the OPA on tripartite entanglement contrasts sharply with its effect on the photon-exciton bipartite entanglement shown in Fig.~\ref{fig:6}. As illustrated in Fig.~\ref{fig:11}, the OPA (green curves) strongly suppresses both the magnitude and the thermal resilience of $R_{\tau}^{\min}$, causing it to vanish at $T\approx 0.05\,\mathrm{K}$. In comparison, the OPA-free case (red curves) maintains the entanglement up to $T\approx 15\,\mathrm{K}$. Therefore, we can conclude that the OPA suppresses the nonreciprocal tripartite entanglement and weakens the tolerance against thermal noise.

\section{\label{level5}Concluding remarks}

In summary, we have investigated nonreciprocal bipartite and tripartite entanglement in a spinning EOMS containing an OPA. The Sagnac-Fizeau effect induced by rotation introduces a directional-dependent frequency shift, while the OPA provides tunable nonlinear gain  and phase, both of which contribute to manipulation of nonreciprocal dynamics. Our results demonstrate that the OPA serves as a versatile tool for controlling nonreciprocal entanglement, with markedly different effects depending on the specific mode pairing for bipartite entanglement.
It is indicated that the OPA can significantly enhance  the photon-exciton entanglement  and  its robustness against the cavity dissipation. In contrast, the OPA suppresses the photon-phonon and exciton-phonon entanglement.  Notably, it is found that the OPA can significantly enhance nonreciprocal tripartite entanglement over a substantially broader detuning range.
We have also studied the influence of temperature on nonreciprocal entanglement. The results show that the thermal robustness of entanglement exhibits strong mode dependence. The OPA can dramatically enhance  the resilience of photon-exciton entanglement against temperature.  Particularly, nonreciprocal photon-exciton entanglement can be generated at room temperature and even higher temperature.  However,  the OPA conversely accelerates the thermal degradation of phonon-involved bipartite and tripartite entanglements.
Our findings establish an effective mechanism for achieving and controlling nonreciprocal entanglement in the EOMS by the use of the OPA. The ability to generate nonreciprocal entanglement and dramatically enhance thermal robustness in selected subsystems opens new possibilities for practical applications in room-temperature quantum information technologies. Future work may explore the extension of these ideas to more complex networks based on the EOMS and the experimental realization of the predicted phenomena using state-of-the-art exciton-polariton platforms.


\begin{acknowledgements}
	
 L.-M. K. is supported by  the Quantum Science and Technology-National Science and Technology Major Project (Grant No. 2024ZD0301000), the Natural Science Foundation of China (NSFC) (Grant Nos. 12247105, 12175060, 12421005), Hunan provincial sci-tech program (Grant No. 2023ZJ1010),  Henan Science and Technology Major Project (No. 241100210400), and  XJ-Lab key project (Grant No. 23XJ02001). H.J. is supported by the National Key R$\&$D Program (GrantNo. 2024YFE0102400), the NSFC (Grants Nos. 11935006 and  12421005). Y.-F.J. is supported by the NSFC (Grant No. 12405029) and the Natural Science Foundation of Henan Province (Grant No. 252300421221).

\end{acknowledgements}

%
	

\begin{thebibliography}{99}

\bibitem{1} R. Horodecki, P Horodecki, M. Horodecki, and K. Horodecki, Quantum entanglement, \href{https://journals.aps.org/rmp/abstract/10.1103/RevModPhys.81.865} {Rev. Mod. Phys. \textbf{81}, 865 (2009)}.

\bibitem{2} A. Einstein, B. Podolsky, and N. Rosen, Can quantum-mechanical description of physical reality be considered complete? \href{https://journals.aps.org/pr/abstract/10.1103/PhysRev.47.777} {Phys. Rev. \textbf{47}, 777 (1935)}.

\bibitem{3} J. S. Bell, On the Einstein Podolsky Rosen paradox, \href{https://journals.aps.org/ppf/abstract/10.1103/PhysicsPhysiqueFizika.1.195} {Physics  \textbf{1}, 195 (1964)}.

\bibitem{4}  J. F. Clauser, M. A. Horne, A. Shimony, and R. A. Holt, Proposed experiment to test local hidden-variable theories, \href{https://journals.aps.org/prl/abstract/10.1103/PhysRevLett.23.880} {Phys. Rev. Lett. \textbf{23}, 880 (1969)}.

\bibitem{5}  S. J. Freedman and J. F. Clauser, Experimental test of local hidden-variable theories, \href{https://journals.aps.org/prl/abstract/10.1103/PhysRevLett.28.938} {Phys. Rev. Lett. \textbf{28}, 938 (1972)}.

\bibitem{6} A. Aspect, P. Grangier, and G. Roger, Experimental tests of realistic local theories via Bell's theorem, \href{https://journals.aps.org/prl/abstract/10.1103/PhysRevLett.47.460} {Phys. Rev. Lett. \textbf{47}, 460 (1981)}.

\bibitem{7}  A. Aspect, J. Dalibard, and G. Roger, Experimental test of Bell's inequalities using time-varying analyzers, \href{https://journals.aps.org/prl/abstract/10.1103/PhysRevLett.49.1804} {Phys. Rev. Lett. \textbf{49}, 1804 (1982)}.

\bibitem{8}  W. Tittel, J. Brendel, B. Gisin, T. Herzog, H. Zbinden, and N. Gisin, Experimental demonstration of quantum correlations over more than 10 km, \href{https://journals.aps.org/pra/abstract/10.1103/PhysRevA.57.3229} {Phys. Rev. A \textbf{57}, 3229 (1998)}.

\bibitem{9}  W. Tittel, J. Brendel, H. Zbinden, and N. Gisin, Violation of Bell inequalities by photons more than 10 km apart, \href{https://journals.aps.org/prl/abstract/10.1103/PhysRevLett.81.3563} {Phys. Rev. Lett. \textbf{81}, 3563 (1998)}.

\bibitem{10} G. Weihs, T. Jennewein, C. Simon, H. Weinfurter, and A. Zeilinger, Violation of Bell's inequality under strict Einstein locality conditions, \href{https://journals.aps.org/prl/abstract/10.1103/PhysRevLett.81.5039} {Phys. Rev. Lett. \textbf{81}, 5039 (1998)}.

\bibitem{11} J.-W. Pan, D. Bouwmeester, M. Daniell, H. Weinfurter, and A. Zeilinger, Experimental test of quantum nonlocality in three-photon Greenberger-Horne-Zeilinger entanglement, \href{https://www.nature.com/articles/35000514} {Nature (London) \textbf{403}, 515 (2000)}.
 R. Schmied, and P. Treutlein, Quantum metrology with nonclassical states of atomic ensembles,  \href{https://journals.aps.org/rmp/abstract/10.1103/RevModPhys.90.035005} {Rev. Mod. Phys. \textbf{90}, 035005 (2018)}.

\bibitem{18}  M. Aspelmeyer, T. J. Kippenberg, and F. Marquardt, Cavity optomechanics, \href{https://journals.aps.org/rmp/abstract/10.1103/RevModPhys.86.1391} {Rev. Mod. Phys. \textbf{86}, 1391 (2014)}.

\bibitem{19}  W. P. Bowen and G. J. Milburn, Quantum Optomechanics (CRC Press, Boca Raton, FL, 2015).

\bibitem{20}  T. J. Kippenberg and K. J. Vahala, Cavity Optomechanics: Back-action at the mesoscale, \href{https://www.science.org/doi/10.1126/science.1156032} {Science \textbf{321}, 1172 (2008)}.

\bibitem{21} G. Huang, A. Beccari, N. J. Engelsen, and T. J. Kippenberg, Room-temperature quantum optomechanics using an ultralow noise cavity, \href{https://www.nature.com/articles/s41586-023-06997-3} {Nature \textbf{626}, 512 (2024)}.

\bibitem{22}  L. Mercier de L$\acute{e}$pinay, C. F. Ockeloen-Korppi, M. J. Woolley, and M. A. Sillanp\"{a}\"{a}, Quantum mechanics-free subsystem with mechanical oscillators, \href{https://www.science.org/doi/10.1126/science.abf2998} {Science \textbf{372}, 625 (2021)}.

\bibitem{23}  E. E.Wollman, C. U. Lei, A. J. Weinstein
\bibitem{12} M. A. Rowe, D. Kielpinski, V. Meyer, C. A. Sackett, W. M. Itano, C. Monroe, and D. J. Wineland, Experimental violation of a Bell's inequality with efficient detection, \href{https://www.nature.com/articles/35057215} {Nature (London) \textbf{409}, 791 (2001)}.

\bibitem{13} M.  A.  Nielsen  and  I.  L.  Chuang,  Quantum  Computation and Quantum Information (Cambridge University Press, Cambridge, 2000).

\bibitem{14} F. J. Duarte and T. Taylor, Quantum Entanglement Engineering and Applications (IOP Press, London, 2021).

\bibitem{15} S. L. Braunstein and P. van Loock, Quantum information with   continuous variables, \href{https://journals.aps.org/rmp/abstract/10.1103/RevModPhys.77.513} {Rev. Mod. Phys. \textbf{77}, 513 (2005)}.

\bibitem{16} F. Flamini, N. Spagnolo, and F. Sciarrino, Photonic quantum information processing: A review, \href{https://iopscience.iop.org/article/10.1088/1361-6633/aad5b2} {Rep. Prog. Phys. \textbf{82}, 016001 (2019)}.

\bibitem{17} L. Pezz\`{e}, A. Smerzi, M. K. Oberthaler,, J. Suh, A. Kronwald, F. Marquardt, A. A. Clerk, and K. C. Schwab, Quantum squeezing of motion in a mechanical resonator, \href{https://www.science.org/doi/10.1126/science.aac5138} {Science \textbf{349}, 952 (2015)}.

\bibitem{24}  T. P. Purdy, P. L. Yu, R. Peterson, N. S. Kampel, and C. A. Regal, Strong optomechanical squeezing of light, \href{https://journals.aps.org/prx/abstract/10.1103/PhysRevX.3.031012} {Phys. Rev. X \textbf{3}, 031012 (2013)}.

\bibitem{25} M. Chegnizadeh, M. Scigliuzzo, A. Youssefi, S. Kono, E. Guzovskii, and T. J. Kippenberg, Quantum collective motion of macroscopic mechanical oscillators, \href{https://www.science.org/doi/10.1126/science.adr8187} {Science \textbf{386}, 1383 (2024)}.

\bibitem{26} S. Kiesewetter, R. Y. Teh, P. D. Drummond, and M. D. Reid, Pulsed entanglement of two optomechanical oscillators and Furry's hypothesis, \href{https://journals.aps.org/prl/abstract/10.1103/PhysRevLett.119.023601} {Phys. Rev. Lett. \textbf{119}, 023601 (2017)}.

\bibitem{27} Y. F. Jiao, S. D. Zhang, Y. L. Zhang, A. Miranowicz, L. M. Kuang, and H. Jing, Nonreciprocal optomechanical entanglement against backscattering losses, \href{https://journals.aps.org/prl/abstract/10.1103/PhysRevLett.125.143605} {Phys. Rev. Lett. \textbf{125}, 143605 (2020)}.

\bibitem{28} Y. F. Jiao,  J. X. Liu, Y. Li, R. Yang, L. M. Kuang, and H. Jing, Nonreciprocal enhancement of remote entanglement between nonidentical mechanical oscillators, \href{https://journals.aps.org/prapplied/abstract/10.1103/PhysRevApplied.18.064008} {Phys. Rev. Applied \textbf{18}, 064008 (2022)}.

\bibitem{29} B. B. Li, L. Ou, Y. Lei, and Y. C. Liu, Cavity optomechanical sensing, \href{https://www.degruyterbrill.com/document/doi/10.1515/nanoph-2021-0256/html} {Nanophotonics \textbf{10}, 2799 (2021)}.

\bibitem{30} S. Forstner, S. Prams, J.Knittel, E. D. van Ooijen, J. D. Swaim, G. I. Harris, A. Szorkovszky, W. P. Bowen, and H. Rubinsztein-Dunlop, Cavity optomechanical magnetometer, \href{https://journals.aps.org/prl/abstract/10.1103/PhysRevLett.108.120801} {Phys. Rev. Lett. \textbf{108}, 120801 (2012)}.

\bibitem{31}  W. J. Westerveld, M. Mahmud-Ul-Hasan, R. Shnaiderman, V. Ntziachristos, X. Rottenberg, S. Severi, and V. Rochus, Sensitive, small, broadband and scalable optomechanical ultrasound sensor in silicon photonics, \href{https://www.nature.com/articles/s41566-021-00776-0} {Nat. Photonics \textbf{15}, 341 (2021)}.

\bibitem{32}  O. Arcizet, P. F. Cohadon, T. Briant, M. Pinard, A. Heidmann, J. M. Mackowski, C. Michel, L. Pinard, O.  Fran\c{c}ais, and L. Rousseau, High-sensitivity optical monitoring of a micromechanical resonator with a quantum-limited optomechanical sensor, \href{https://journals.aps.org/prl/abstract/10.1103/PhysRevLett.97.133601} {Phys. Rev. Lett. \textbf{97}, 133601 (2006)}.

\bibitem{33}  I. Wilson-Rae, N. Nooshi, W. Zwerger, and T. J. Kippenberg, Theory of ground state cooling of a mechanical oscillator using  dynamical backaction, \href{https://journals.aps.org/prl/abstract/10.1103/PhysRevLett.99.093901} {Phys. Rev. Lett. \textbf{99}, 093901 (2007)}.

\bibitem{34} F. Marquardt, J. P. Chen, A. A. Clerk, and S. M. Girvin, Quantum theory of cavity-assisted sideband cooling of mechanical motion, \href{https://journals.aps.org/prl/abstract/10.1103/PhysRevLett.99.093902} {Phys. Rev. Lett. \textbf{99}, 093902 (2007)}.

\bibitem{35}  J. Chan, T. P. Alegre, A. H. Safavi-Naeini, J. T. Hill, A. Krause, S. Groeblacher, M. Aspelmeyer, and O. Painter, Laser cooling of a nanomechanical oscillator into its quantum ground state, \href{https://www.nature.com/articles/nature10461} {Nature (London) \textbf{478}, 89 (2011)}.

\bibitem{36}   J. D. Teufel, T. Dormer, D. Li, J. W. Harlow, M. S. Allman, K. Cicak, A. J. Sirois, J. D. Whittaker, K. W. Lehnert, and R. W. Simmonds, Sideband cooling of micromechanical motion to the quantum ground state, \href{https://www.nature.com/articles/nature10261} {Nature (London) \textbf{475}, 359 (2011)}.

\bibitem{37}   M. Paternostro, D. Vitali, S. Gigan, M. S. Kim, C. Brukner, J. Eisert, and M. Aspelmeyer, Creating and probing multipartite macroscopic entanglement with light, \href{https://journals.aps.org/prl/abstract/10.1103/PhysRevLett.99.250401} {Phys. Rev. Lett. \textbf{99}, 250401 (2007)}.

\bibitem{38}   M. J. Hartmann and M. B.Plenio, Steady state entanglement in the mechanical vibrations of two dielectric membranes, \href{https://journals.aps.org/prl/abstract/10.1103/PhysRevLett.101.200503} {Phys. Rev. Lett. \textbf{101}, 200503 (2008)}.

\bibitem{39}  S. Mancini, V. Giovannetti, D. Vitali, and P. Tombesi, Entangling macroscopic oscillators exploiting radiation pressure, \href{https://journals.aps.org/prl/abstract/10.1103/PhysRevLett.88.120401} {Phys. Rev. Lett. \textbf{88}, 120401 (2002)}.

\bibitem{40}   D. Vitali, S. Gigan, A. Ferreira, H. R. Bohm, P. Tombesi, A. Guerreiro, V. Vedral, A. Zeilinger, and M. Aspelmeyer, Optomechanical entanglement between a movable mirror and a  cavity field, \href{https://journals.aps.org/prl/abstract/10.1103/PhysRevLett.98.030405} {Phys. Rev. Lett. \textbf{98}, 030405 (2007)}.

\bibitem{41}   C. Genes, D. Vitali, and P. Tombesi, Simultaneous cooling and entanglement of mechanical modes of a micromirror in an  optical cavity, \href{https://iopscience.iop.org/article/10.1088/1367-2630/10/9/095009/pdf} {New J. Phys. \textbf{10}, 095009 (2008)}.

\bibitem{42}  L. Tian, Robust photon entanglement via quantum interference in optomechanical interfaces, \href{https://journals.aps.org/prl/abstract/10.1103/PhysRevLett.110.233602} {Phys. Rev. Lett. \textbf{110}, 233602 (2013)}.

\bibitem{43}  S. Barzanjeh, E. S. Redchenko, M. Peruzzo, M. Wulf, D. P. Lewis, G. Arnold, and J. M. Fink, Stationary entangled radianon from micromechanical motion, \href{https://www.nature.com/articles/s41586-019-1320-2} {Nature (London) \textbf{570}, 480 (2019)}.

\bibitem{44}   Y. D. Wang and A. A. Clerk, Reservoir-engineered entanglement in optomechanical systems, \href{https://journals.aps.org/prl/abstract/10.1103/PhysRevLett.110.253601} {Phys. Rev. Lett. \textbf{110}, 253601 (2013)}.

\bibitem{45}   T. A. Palomaki, J. D. Teufel, R. W. Simmonds, and K. W. Lehnert, Entangling mechanical motion with microwave fields, \href{https://www.science.org/doi/10.1126/science.1244563} {Science \textbf{342}, 710 (2013)}.

\bibitem{46}  J. Q. Liao, Q. Q. Wu, and F. Nori, Entangling two macroscopic mechanical mirrors in a two-cavity optomechanical system, \href{https://journals.aps.org/pra/abstract/10.1103/PhysRevA.89.014302} {Phys. Rev. A \textbf{89}, 014302 (2014)}.

\bibitem{47}  D. G. Lai, J. Q. Liao, A. Miranowicz, and F. Nori, Noise-tolerant optomechanical entanglement via synthetic magnetism, \href{https://journals.aps.org/prl/abstract/10.1103/PhysRevLett.129.063602} {Phys. Rev. Lett. \textbf{129}, 063602 (2022)}.

\bibitem{48}   J. Huang, D. G. Lai, and J. Q. Liao, Thermal-noise-resistant optomechanical entanglement via general dark-mode control,  \href{https://journals.aps.org/pra/abstract/10.1103/PhysRevA.106.063506} {Phys. Rev. A \textbf{106}, 063506 (2022)}.

\bibitem{49}   J. X. Liu, Y. F. Jiao, Y  Li, X. W. Xu, Q. Y. He, and H. Jing, Phase-controlled asymmetric optomechanical entanglement against optical backscattering, \href{https://link.springer.com/article/10.1007/s11433-022-2043-3} {Sci. China: Phys. Mech.  Astron. \textbf{66}, 230312 (2023)}.

\bibitem{50} C. F. Ockeloen-Korppi, E. Damsk\"{a}gg, J. M. Pirkkalainen, M. Asjad, A. A. Clerk, F. Massel, M. J. Woolley, and M. A. Sillanp\"{a}\"{a}, Stabilized entanglement of massive mechanical oscillators, \href{https://www.nature.com/articles/s41586-018-0038-x} {Nature (London) \textbf{556}, 478 (2018)}.

\bibitem{51}   R. Riedinger, A. Wallucks, I. Marinkovic, C. Loschnauer, M. Aspelmeyer, S. Hong, and S. Groblacher, Remote quantum entanglement between two micromechanical oscillators, \href{https://www.nature.com/articles/s41586-018-0036-z} {Nature (London) \textbf{556}, 473 (2018)}.

\bibitem{52}   L. Mercier de L?inay, C. F. Ockeloen-Korppi, M. J. Woolley, and M. A. Sillanpaa, Quantum mechanics-free subsystem with  mechanical oscillators, \href{https://www.science.org/doi/full/10.1126/science.abf5389} {Science \textbf{372}, 625 (2021)}.

\bibitem{53}   S. Kotler, G. A. Peterson, E. Shojaee, F. Lecocq, K. Cicak, A.  Kwiatkowski,  S.  Geller,  S.  Glancy,  E.  Knill,  R.  W. Simmonds, J. Aumentado, and J. D. Teufel, Direct observation of deterministic macroscopic entanglement, \href{https://www.science.org/doi/10.1126/science.abf2998} {Science \textbf{372}, 622  (2021)}.

\bibitem{54}  J. Kasprzak,M. Richard, S. Kundermann, A. Baas, P. Jeambrun, J. M. J. Keeling, F. M. Marchetti, M. H. Szyma\'nska, R. Andr\'e, J. L. Staehli, V. Savona, P. B. Littlewood, B. Deveaud, and L. S. Dang, Bose-Einstein condensation of exciton polaritons, \href{https://www.nature.com/articles/nature05131} {Nature (London) \textbf{443}, 409 (2006)}.

\bibitem{55}  D. Ballarini, M. De Giorgi, E. Cancellieri, R. Houdr\'e, E. Giacobino, R. Cingolani, A. Bramati, G. Gigli, and D. Sanvitto, All-optical polariton transistor, \href{https://www.nature.com/articles/ncomms2734} {Nat. Commun. \textbf{4}, 1778 (2013)}.

\bibitem{56}   B. Jusserand, A. N. Poddubny, A. V. Poshakinskiy, A. Fainstein, and A. Lemaitre, Polariton resonances for ultrastrong coupling cavity optomechanics in GaAs/AlAs multiple quantum wells, \href{https://journals.aps.org/prl/abstract/10.1103/PhysRevLett.115.267402} {Phys. Rev. Lett. \textbf{115}, 267402 (2015)}.

\bibitem{57}   A. Fainstein, N. D. Lanzillotti-Kimura, B. Jusserand, and B. Perrin, Strong optical-mechanical coupling in a vertical GaAs/AlAs microcavity for subterahertz phonons and near-infrared light, \href{https://journals.aps.org/prl/abstract/10.1103/PhysRevLett.110.037403} {Phys. Rev. Lett. \textbf{110}, 037403 (2013)}.

\bibitem{58}   R. de Oliveira, M. Colombano, F. Malabat, M. Morassi, A. Lema\^{i}tre, and I. Favero, Whispering-gallery quantum-well exciton polaritons in an indium gallium arsenide microdisk cavity, \href{https://journals.aps.org/prl/abstract/10.1103/PhysRevLett.132.126901} {Phys. Rev. Lett. \textbf{132}, 126901 (2024)}.

\bibitem{59}   B. Guha, F. Marsault, F. Cadiz, L. Morgenroth, V. Ulin, V. Berkovitz, A. Lema\^{i}tre, C. Gomez, A. Amo, S. Combri\'e, B. G\'{e}rard, G. Leo, and I. Favero, Surface-enhanced gallium arsenide photonic resonator with quality factor of $6\times 10^6$, \href{https://www.researchgate.net/publication/301819264_Surface-enhanced_gallium_arsenide_photonic_resonator_with_quality_factor_of_6_10} {Optica \textbf{4}, 218 (2017)}.

\bibitem{60}  N. Carlon Zambon, Z. Denis, R. De Oliveira, S. Ravets, C. Ciuti, I. Favero, and J. Bloch, Enhanced cavity optomechanics with quantum-well exciton polaritons, \href{https://journals.aps.org/prl/abstract/10.1103/PhysRevLett.129.093603} {Phys. Rev. Lett. \textbf{129}, 093603 (2022)}.

\bibitem{61}   R. Su, S. Ghosh, J. Wang, S. Liu, C. Diederichs, T. C. H. Liew, and Q. Xiong, Observation of exciton polariton condensation in a perovskite lattice at room temperature, \href{https://www.nature.com/articles/s41567-019-0764-5} {Nat. Phys. \textbf{16}, 301 (2020)}.

\bibitem{62}   S. Christopoulos, G. Baldassarri H\"{o}ger von H\"{o}gersthal, A. J. D. Grundy, P. G. Lagoudakis, A. V. Kavokin, J. J. Baumberg, G. Christmann, R. Butt\'e, E. Feltin, J.-F. Carlin, and N. Grandjean, Room-temperature polariton lasing in semiconductor microcavities, \href{https://journals.aps.org/prl/abstract/10.1103/PhysRevLett.98.126405} {Phys. Rev. Lett. \textbf{98}, 126405 (2007)}.

\bibitem{63}  A. Amo, J. Lefr\`{e}re, S. Pigeon, C. Adrados, C. Ciuti, I. Carusotto, R. Houdr\'e, E. Giacobino, and A. Bramati, Superfluidity of polaritons in semiconductor microcavities, \href{https://www.nature.com/articles/nphys1364} {Nat. Phys. \textbf{5}, 805 (2009)}.

\bibitem{64}  O. Kyriienko, T. C. H. Liew, and I. A. Shelykh, Optomechanics with cavity polaritons: Dissipative coupling and unconventional bistability, \href{https://journals.aps.org/prl/abstract/10.1103/PhysRevLett.112.076402} {Phys. Rev. Lett. \textbf{112}, 076402 (2014)}.

\bibitem{65}  D. Bajoni, E. Semenova, A. Lema\^{i}tre, S. Bouchoule, E. Wertz, P. Senellart, S. Barbay, R. Kuszelewicz, and J. Bloch, Optical bistability in a GaAs-based polariton diode, \href{https://journals.aps.org/prl/abstract/10.1103/PhysRevLett.101.266402} {Phys. Rev. Lett. \textbf{101}, 266402 (2008)}.

\bibitem{66}   Z.-F. Yu, J.-K. Xue, L. Zhuang, J. Zhao, and W.-M. Liu, Non-Hermitian spectrum and multistability in exciton-polariton condensates, \href{https://journals.aps.org/prb/abstract/10.1103/PhysRevB.104.235408} {Phys. Rev. B \textbf{104}, 235408 (2021)}.

\bibitem{67}   K. C. Yellapragada, N. Pramanik, S. Singh, and P. A. Lakshmi, Optomechanical effects in a macroscopic hybrid system, \href{https://journals.aps.org/pra/abstract/10.1103/PhysRevA.98.053822} {Phys. Rev. A \textbf{98}, 053822 (2018)}.

\bibitem{68}   Z.-F. Yu and J.-K. Xue, Photonic transistor based on a coupledcavity system with polaritons, \href{https://www.researchgate.net/publication/372466268_Photonic_transistor_based_on_a_coupled-cavity_system_with_polaritons} {Opt. Express \textbf{31}, 26276 (2023)}.

\bibitem{69}   Z.-F. Yu, P.-F. Yan, J.-M. Gao, F.-Q. Hu, and J.-K. Xue, Photonic negative differential transistor based on cavity polaritons, \href{https://iopscience.iop.org/article/10.1088/1367-2630/acfd54} {New J. Phys. \textbf{25}, 103009 (2023)}.

\bibitem{70}   Y. V. Kartashov and D. V. Skryabin, Two-dimensional topological polariton laser, \href{https://journals.aps.org/prl/abstract/10.1103/PhysRevLett.122.083902} {Phys. Rev. Lett. \textbf{122}, 083902 (2019)}.

\bibitem{71}   A. Amo, T. C. H. Liew, C. Adrados, R. Houdr\'e, E. Giacobino, A. V. Kavokin, and A. Bramati, Exciton-polariton spin switches, \href{https://www.nature.com/articles/nphoton.2010.79} {Nat. Photonics \textbf{4}, 361 (2010)}.

\bibitem{72}   R. Cerna, Y. L\'eger, T. K. Para\"{i}so, M. Wouters, F. Morier-Genoud, M. T. Portella-Oberli, and B. Deveaud, Ultrafast tristable spin memory of a coherent polariton gas, \href{https://www.nature.com/articles/ncomms3008} {Nat. Commun. \textbf{4}, 2008 (2013)}.

\bibitem{73}  T. Gao, P. S. Eldridge, T. C. H. Liew, S. I. Tsintzos, G. Stavrinidis, G. Deligeorgis, Z. Hatzopoulos, and P. G. Savvidis, Polariton condensate transistor switch, \href{https://journals.aps.org/prb/abstract/10.1103/PhysRevB.85.235102} {Phys. Rev. B \textbf{85}, 235102 (2012)}.

\bibitem{74}  Z.-F. Yu, P.-F. Yan, J.-M. Gao, F.-Q. Hu, Z. Zhang, A.-X. Zhang, and J.-K. Xue, Nonreciprocal photonic transistor with a spinning polaritonic microcavity, \href{https://journals.aps.org/pra/abstract/10.1103/PhysRevA.111.013517} {Phys. Rev. A \textbf{111}, 013517 (2025)}.

\bibitem{75}     H. S. Nguyen, D. Vishnevsky, C. Sturm, D. Tanese, D. Solnyshkov, E. Galopin, A. Lema\^{i}tre, I. Sagnes, A. Amo, G. Malpuech, and J. Bloch, Realization of a double-barrier resonant tunneling diode for cavity polaritons, \href{https://journals.aps.org/prl/abstract/10.1103/PhysRevLett.110.236601} {Phys. Rev. Lett. \textbf{110}, 236601 (2013)}.

\bibitem{76}   X. Zuo, Z.-Y. Fan, H. Qian, and J. Li, Entangling excitons with microcavity photons, \href{https://journals.aps.org/prresearch/abstract/10.1103/PhysRevResearch.6.023089} {Phys. Rev. Res. \textbf{6}, 023089 (2024)}.

\bibitem{76-1}   J. Huang and Z. Zhang, Room-temperature exciton-vibration-photon entanglement in polariton optomechanics,  \href{https://journals.aps.org/pra/abstract/10.1103/2dfx-23qp} {Phys. Rev. A \textbf{112}, 013509 (2025)}.

\bibitem{77}  G.  Adesso and F. Illuminati,  Continuous variable tangle, monogamy inequality, and entanglement sharing in Gaussian states of continuous variable systems, \href{https://iopscience.iop.org/article/10.1088/1367-2630/8/1/015} {New J. Phys. \textbf{8},  15 (2006)}.

\bibitem{78} G. Adesso and F. Illuminati, Entanglement in continuous-variable systems: Recent advances and current perspectives, \href{https://iopscience.iop.org/article/10.1088/1751-8113/40/28/S01} {J. Phys. A \textbf{40}, 7821 (2007)}.

\end{thebibliography}
\end{document}